\PassOptionsToPackage{final}{graphicx}
\RequirePackage{ifthen}
\documentclass[final,a4paper,USenglish,cleveref,numberwithinsect,autoref]{lipics-v2021}
\pdfoutput=1 %

\hideLIPIcs %

\newboolean{longversion}
\setboolean{longversion}{true}
\newboolean{anonymous}
\setboolean{anonymous}{false}
\newboolean{spellingmode}
\setboolean{spellingmode}{false}
\makeatletter
\@ifclasswith{lipics-v2021}{anonymous}{%
	\setboolean{anonymous}{true}
}{%
}
\@ifclasswith{lipics-v2021}{review}{%
	\setboolean{longversion}{false}
}{
	\nolinenumbers
}
\makeatother
\ifthenelse{\boolean{spellingmode}}{%
	\nolinenumbers
	\AtEndPreamble{
	\hyphenpenalty=50000 %
	\pagestyle{empty}%
	\thispagestyle{empty}%
	\def\pagestyle#1{}%
	\def\thispagestyle#1{}%
	\def\labelmarginpar#1{}%
	}
}{}

\usepackage[utf8]{inputenc}
\usepackage[T1]{fontenc}

\usepackage{proof}

\usepackage{algorithm, algpseudocode}
\newcommand{\code}[1]{\texttt{#1}}
\usepackage{marvosym}
\usepackage{enumitem}
\usepackage{listings}
\usepackage[obeyDraft]{todonotes}
\usepackage{url}
\usepackage{subcaption}
\usepackage{svg}
\usepackage{pdfpages}
\usepackage{booktabs}
\usepackage{longtable}
\usepackage{placeins}
\usepackage{wrapfig}
\usepackage{etex}
\usepackage{etoolbox}
\usepackage{environ}
\usepackage{ifdraft}
\usepackage[outline]{contour}
\contourlength{1.2pt}

\usepackage{hyperref}
\usepackage{twshowkeys}

\usepackage{tikz}
\usetikzlibrary{patterns, intersections, positioning, shapes.geometric, calc,
	backgrounds, fit,
	automata, arrows, decorations.pathreplacing,decorations.pathmorphing}
\usetikzlibrary{cd}

\hypersetup{
  final,
  hidelinks,
}

\theoremstyle{plain}
\theoremstyle{definition}
\newtheorem{notation}[theorem]{Notation}
\newtheorem{assumption}[theorem]{Assumption}

\newcommand{\twnote}[1]{\todo{TW: #1}}
\newcommand{\fvnote}[1]{\todo{FV: #1}}
\setuptodonotes{
  backgroundcolor=yellow!30,
  size=\tiny,
  noline,
}

\makeatletter
\newsavebox{\RuleLabelTag}
\newcommand\rulelabel[2]{{%
	\edef\@currentlabel{#2}%
	\ltx@label{{#1}}%
	\sbox{\RuleLabelTag}{\text{\hspace*{4pt}\upshape (#2)}}%
	\usebox{\RuleLabelTag}%
}}
\makeatother

\newcommand{\M}{{\mathcal{M}}}
\newcommand{\N}{{\mathcal{N}}}
\newcommand{\Nat}{{\mathbb{N}}}
\newcommand{\R}{{\mathcal{R}}}
\def\S{{\mathcal{S}}}

\newcommand{\LTS}{\ensuremath{\mathsf{LTS}}}
\newcommand{\tuple}[1]{\ensuremath{{\langle #1 \rangle}}}
\newcommand{\ascii}[1]{\text{\upshape\texttt{#1}}}

\newcommand{\trace}{{\mathit{trace}}}
\newcommand{\obs}{\trace}

\newcommand{\Code}{{\mathsf{Code}}}

\newcommand{\partialto}{{\rightharpoonup}}
\newcommand{\dom}{{\mathsf{dom}}}
\newcommand{\SI}{{\mathrel{\mathcal{I}}}}
\renewcommand{\Im}{{\mathsf{im}}}
\renewcommand{\rho}{\varrho}
\renewcommand{\epsilon}{\varepsilon}
\newcommand{\textqt}[1]{`#1'}
\newcommand{\set}[2][]{%
  \ifthenelse{\equal{#2}{}}{%
    \ensuremath{{#1\emptyset}}%
  }{%
    \ensuremath{{#1\{#2#1\}}}%
  }%
}

\newcommand{\imgcoffee}{\includegraphics[scale=0.2]{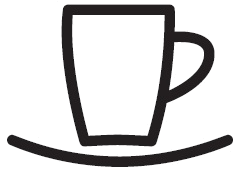}}
\newcommand{\imgespresso}{\includegraphics[scale=0.15]{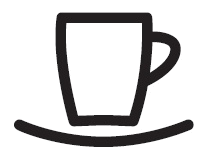}}

\newcommand{\takeout}[1]{\relax}
\NewEnviron{iflong}{%
	\ifthenelse{\boolean{longversion}}{%
		\todo{Only in the long version}
		\BODY%
	}{%
	}%
}

\makeatletter
\newcommand{\printcoqreferences}{%
	\ifthenelse{\value{coqreferencecounter}=0}{%
		\textit{No coq references yet.}%
	}{%
	\begin{description}%
	\coqreferencesbuffer
	\end{description}%
	}%
}

\definecolor[named]{lipicsBulletGrayPreamble}{rgb}{0.60,0.60,0.61}
\newsavebox{\coqiconbox}
\sbox{\coqiconbox}{%
\raisebox{-2pt}[4pt][0pt]{%
\includegraphics[height=12pt]{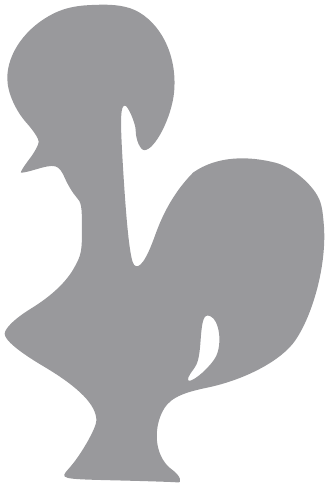}%
}%
}%

\def\coqicon{%
	\hyperref[coqtoc]{\usebox{\coqiconbox}}%
}

\DeclareUrlCommand\UScore{\urlstyle{tt}}
\newcounter{coqreferencecounter}
\def\coqreferencesbuffer{}
\newcommand{\auxcoqref}[4]{%
	\stepcounter{coqreferencecounter}%
	\g@addto@macro\coqreferencesbuffer{%
		\item[\autoref{#1}] $\hat{=}$ \UScore{#3} (in \UScore{#2})%
		\ifstrempty{#4}{.}{
		-- #4%
		}%
	}%
}
\newcommand{\writeauxcoqref}[4]{%
	\write\@auxout{%
		\noexpand\auxcoqref%
			{\unexpanded{#1}}%
			{\unexpanded{#2}}%
			{\unexpanded{#3}}%
			{\unexpanded{#4}}%
	}%
}

\newcounter{auxcoqlabelcounter}
\newcommand{\coqref}[3]{%
	\stepcounter{auxcoqlabelcounter}%
	\edef\auxcoqcurrentlabel{auxcoqlabel\arabic{auxcoqlabelcounter}}%
	\ifcsname TWSK@oldlabel\endcsname%
		\TWSK@oldlabel{\auxcoqcurrentlabel}%
	\else%
		\expandafter\label{\auxcoqcurrentlabel}%
	\fi%
	\coqicon%
	\expandafter\writeauxcoqref\expandafter{\auxcoqcurrentlabel}{#1}{#2}{#3}%
}
\makeatother

\makeatletter
\newcommand{\proofappendixbegin}[2]{%
  \phantomsection%
  \def\proofappendix@qedsymbolmissing{\qed}%
  \subsection*{{#1~\autoref{#2}}}%
  \def\qedhere{\global\def\proofappendix@qedsymbolmissing{}\qed}%
  \addcontentsline{toc}{subsection}{#1~\autoref{#2}}%
  \label{#2:proof}%
}
\newcommand{\proofappendixend}{%
  \proofappendix@qedsymbolmissing%
  \let\qedhere\oldqedhere%
}
\let\oldqedhere\qedhere
\makeatother

\newenvironment{proofappendix}[2][Proof of]{%
  \proofappendixbegin{#1}{#2}%
}{%
  \proofappendixend%
  \par%
}

\newcommand{\hidestatename}[1]{}
\colorlet{transitioncolor}{black}
\tikzset{
  lts/.style={
    auto,
    node distance=2.5cm,
    state/.style={
      shape=ellipse,
      outer sep=2pt,
      inner xsep=0pt,
      inner ysep=2pt,
      fill=none,
      draw=lipicsYellow,
      line width=1pt,
      minimum size=9mm,
      text depth=0mm,
    },
    every loop/.style={looseness=10, min distance=4mm},
    transition/.style={
      line width=.7pt,
      -{Triangle[length=4pt,width=4pt]},
      draw=transitioncolor,
      every node/.append style={
        font=\upshape\footnotesize,
        inner sep=2pt,
        },
    },
		generalized transition/.style={
			transition,
			double distance=1pt,
		},
    every initial by arrow/.style={
      transition,
      shorten <=.5pt,
      every node/.append style={
        overlay,
      },
    },
    initial text={},
    initial distance=4mm,
    frame decoration/.style={
      line width=1pt,
      draw=black!30,
      rounded corners=2pt,
    },
    white glow/.style={
      preaction={draw=white,line width=3pt,},
    },
    white background node/.style={
      shape=rectangle,
      fill=white,
      inner sep=1pt,
      outer sep=3pt,
      rounded corners=1pt,
    },
  },
	empty states/.style={
		state/.append style={
			minimum size=2mm,
			fill=lipicsYellow,
		},
	},
	realm border/.style={
		draw=black!40,
		rounded corners=2pt,
		line width=1pt,
	},
	realm reserve space for title/.style={
		inner ysep=4mm,
		yshift=2mm, %
	},
	realm title/.style={
		draw=none,
		inner sep=3pt,
		fill=black!40,
		text=white,
		font=\scriptsize,
		rounded corners=2pt,
	},
}

\newcommand{\drawRealm}[2]{%
	\begin{scope}[on background layer]
		\node[realm border, realm reserve space for title, fit=#2] (realm border) {};
		\node[realm title] at (realm border.north) {\ensuremath{#1}};
	\end{scope}
}

\newcommand{\drawLtsFrame}{%
	\begin{scope}[on background layer]
		\node[realm border,fit=(current bounding box)] {};
	\end{scope}
}

\newcommand{\drawTransto}[1]{%
    \node[font=\scriptsize,inner sep=1mm,minimum width=12pt,inner ysep=1pt] (label) {\ensuremath{#1}} ;
    \node[outer sep=0pt,anchor=west] (a) at (label.south west) {\phantom{X}};
    \node[outer sep=0pt,anchor=east] (b) at ([xshift=2pt]label.south east) {};
    \coordinate (arrow tip) at (b.east |- a.west);
    \draw[transition] (a.west) to (arrow tip);
}

\newcommand{\xtransto}[2][]{%
\mathrel{%
  \begin{tikzpicture}[baseline=(a.base), lts, inner sep=0pt]
    \drawTransto{#2}
  \end{tikzpicture}%
		\ifthenelse{\equal{#1}{}}{%
		}{%
			\ensuremath{{}_{#1}}%
		}%
}%
}

\newcommand{\xTransto}[2][]{%
\mathrel{%
  \begin{tikzpicture}[baseline=(a.base), lts, inner sep=0pt]
    \node[font=\scriptsize,inner sep=1mm,minimum width=12pt,inner ysep=1pt] (label) {\ensuremath{#2}} ;
    \node[outer sep=0pt,anchor=west] (a) at ([yshift=-1pt]label.south west) {\phantom{X}};
    \node[outer sep=0pt,anchor=east] (b) at ([xshift=2pt]label.east |- a.south) {};
    \coordinate (arrow tip) at (b.east |- a.west);
    \draw[generalized transition] (a.west) to (arrow tip);
  \end{tikzpicture}%
		\ifthenelse{\equal{#1}{}}{%
		}{%
			\ensuremath{{}_{#1}}%
		}%
}%
}

\newcommand{\transto}[1][]{
\mathord{%
  \xtransto[#1]{}%
}%
}

\newcommand{\notransto}[1][]{
\mathrel{%
  \begin{tikzpicture}[baseline=(a.base), lts, inner sep=0pt]
    \drawTransto{}
    \coordinate (arrow mid) at (label.south |- arrow tip);
    \coordinate (strike through line) at (1pt,3pt);
    \draw[transition,-] ($ (arrow mid) - (strike through line) $) --
    ($ (arrow mid) + (strike through line) $);
  \end{tikzpicture}%
}%
}

\title{Action Codes}
\ifthenelse{\boolean{anonymous}}{
\author{Anonymous}{Anonymous affil.}{}{}{}
}{
\author{Frits Vaandrager}{Radboud University, Nijmegen, the Netherlands\and\url{https://www.cs.ru.nl/F.Vaandrager/}}{F.Vaandrager@cs.ru.nl}{https://orcid.org/0000-0003-3955-1910}{Supported by the NWO TOP project 612.001.852}
\author{Thorsten Wißmann}{Radboud University, Nijmegen, the Netherlands\and Friedrich-Alexander-Universität Erlangen-Nürnberg, Germany\and\url{https://thorsten-wissmann.de}}{T.Wissmann@cs.ru.nl}{https://orcid.org/0000-0001-8993-6486}{Supported by the NWO TOP project 612.001.852}
}

\ifthenelse{\boolean{anonymous}}{}{
\acknowledgements{%
As part of an MSc thesis project under supervision of the first author, Timo Maarse studied a different and more restricted type of action codes (called action refinements) \cite{Maa20}.
It turned out, however, that for these action codes, the concretization operator is not monotone.  The present paper arose from our efforts to fix this problem.
We thank Paul Fiter\u{a}u-Bro\c{s}tean for examples of the use of action codes in model learning.
We thank Jules Jacobs for helpful discussions about Coq.
The first author would like to thank Rocco De Nicola for his hospitality at IMT Lucca during the work on this paper.
}
}

\authorrunning{F.\ Vaandrager and T.\ Wißmann}

\Copyright{F.\ Vaandrager and T.\ Wißmann}

\ccsdesc[500]{Theory of computation}

\keywords{Automata, Models of Reactive Systems, LTS, Action Codes, Action Refinement, Action Contraction, Galois Connection, Model-Based Testing, Model Learning} %

\category{} %

\ifthenelse{\boolean{anonymous}}{}{
\relatedversion{} %
\relatedversiondetails{Full Version With Appendix}{https://arxiv.org/abs/2301.00199}
}

\begin{document}
\maketitle

\begin{abstract}
  We provide a new perspective on the problem how high-level state machine models with abstract actions can be related to low-level models in which these actions are refined by sequences of concrete actions. We describe the connection between high-level and low-level actions using \emph{action codes}, a variation of the prefix codes known from coding theory.
  For each action code $\R$, we introduce a \emph{contraction} operator $\alpha_\R$ that turns a low-level model $\M$ into a high-level model, and a \emph{refinement} operator $\rho_\R$ that transforms a high-level model $\N$ into a low-level model.
  We establish a Galois connection
  $\rho_\R(\N) \sqsubseteq \M  \Leftrightarrow \N \sqsubseteq \alpha_\R(\M)$,
  where $\sqsubseteq$ is the well-known simulation preorder.
  For conformance, we typically want to obtain an overapproximation of model $\M$. To this end, we also introduce a \emph{concretization} operator $\gamma_\R$, which behaves like the refinement operator but adds arbitrary behavior at intermediate points, giving us a second Galois connection
  $\alpha_\R(\M) \sqsubseteq \N \Leftrightarrow \M \sqsubseteq \gamma_\R(\N)$.
  Action codes may be used to construct adaptors that translate between concrete and abstract actions during learning and testing of Mealy machines.
  If Mealy machine $\M$ models a black-box system then $\alpha_\R(\M)$ describes the behavior that can be observed by a learner/tester that interacts with this system via an adaptor derived from code $\R$.
  Whenever $\alpha_\R(\M)$ implements (or conforms to) $\N$, we may conclude that $\M$ implements (or conforms to) $\gamma_{\R} (\N)$.
\end{abstract}

\section{Introduction}

Labeled transition systems (LTSs) constitute one of the most fundamental modeling mechanisms in Computer Science.
An LTS is a rooted, directed graph whose nodes represent \emph{states} and whose edges are labeled with \emph{actions} and represent \emph{state transitions}.
LTS-based formalisms such as Finite Automata \cite{HU79}, Finite State Machines \cite{LeeY96}, I/O automata \cite{Ly96}, IOTSs \cite{Tret96}, and process algebras \cite{HandbookPA} have been widely used to model and analyze a broad variety of reactive systems, and a rich body of theory has been developed for them.

In order to manage the complexity of computer-based systems, designers structure such systems into hierarchical layers. This allows them to describe and analyze systems at different levels of abstraction.  Many LTS-based frameworks have been proposed to formally relate models at different hierarchical levels, e.g. \cite{HandbookPA,Garavel22,LT87,Ya00}.
In most of these frameworks, the states of a high-level LTS correspond to sets of states of a low-level LTS via simulation or bisimulation-like relations. However, the actions are fixed and considered to be atomic. Actions used at a lower level of abstraction can be hidden at a higher level, but higher-level actions will always be available at the lower level.
For this reason, Rensink \& Gorrieri \cite{GorrieriR01,RensinkG01} argue that these (bi)simulations  relate systems at the same conceptual level of abstraction, and therefore they call them \emph{horizontal} implementation relations.
They contrast them with \emph{vertical} implementation relations that compare systems that belong to conceptually different abstraction levels, and have different alphabets of actions.

A prototypical example of a hierarchical design is a computer network. To reduce design complexity, such a network is organized as a stack of layers or levels, each one built upon the one below it \cite{TanenbaumW11}. Examples are the transport layer, with protocols such as TCP and UDP, and the physical layer, concerned with transmitting raw bits over a communication channel.
Now consider a host that receives an TCP packet in some state $s$. If $P$ is the set of possible packets then, in an LTS model of the transport layer, state $s$ will contain outgoing transitions labeled with action $\mathit{receive}(p)$, for each $p \in P$. At the physical layer, however, receipt of a packet corresponds to a sequence of $\mathit{receive}(b)$ actions, with $b$ a bit in $\{0, 1\}$.  Only after the final bits have arrived, the host knows which packet was actually received.
Mechanisms for transforming high-level actions into sequences (or processes) of low-level actions have been addressed extensively in work on action refinements \cite{GorrieriR01}.  These approaches, however, are unable to describe the above scenario in a satisfactory manner and somehow assume that a host upfront correctly guesses the packet that it will receive, even before the first bit has arrived. In order to illustrate this problem, we consider the simplified example of an LTS with a distinguished initial state, displayed in Figure~\ref{ab}, which accepts either input $a$ or input $b$.
\begin{figure}[t]%
	\begin{hideshowkeys}%
	\begin{subfigure}[b]{.21\textwidth}%
	\begin{tikzpicture}[lts,empty states,node distance=1.2cm,
	]
	\node[state, initial] (q0) {};
	\node[state,above right of=q0] (q1) {};
	\node[state,below right of=q0] (q2) {};
	\draw[transition]
	(q0) edge node{$a$} (q1)
	(q0) edge node[swap]{$b$} (q2)
	;
	\end{tikzpicture}%
	\caption{Original System}%
	\label{ab}%
	\end{subfigure}%
	\hfill%
	\begin{subfigure}[b]{.36\textwidth}%
		\begin{tikzpicture}[lts,empty states,node distance=1.2cm]
		\node[state, initial] (q0) {};
		\node[state,above right of=q0] (q1) {};
		\node[state,below right of=q0] (q2) {};
		\node[state,right of=q1] (q3) {};
		\node[state,right of=q2] (q4) {};
		\node[state,right of=q3] (q5) {};
		\node[state,right of=q4] (q6) {};
		\draw[transition]
		(q0) edge node{$1$} (q1)
		(q0) edge node[swap]{$1$} (q2)
		(q1) edge node{$4$} (q3)
		(q2) edge node{$4$} (q4)
		(q3) edge node{$1$} (q5)
		(q4) edge node{$2$} (q6)
		;
		\end{tikzpicture}
		\caption{Existing action refinements}%
		\label{fig:existing refinement}%
	\end{subfigure}%
	\hfill%
	\begin{subfigure}[b]{.39\textwidth}%
		\begin{tikzpicture}[lts,empty states,node distance=1.2cm]
		\node[state, initial] (q0) {};
		\node[state,right of=q0] (q1) {};
		\node[state,right of=q1] (q3) {};
		\node[state,above right of=q3] (q5) {};
		\node[state,below right of=q3] (q6) {};
		\draw[transition]
		(q0) edge node{$1$} (q1)
		(q1) edge node{$4$} (q3)
		(q3) edge node{$1$} (q5)
		(q3) edge node[swap]{$2$} (q6)
		;
		\end{tikzpicture}%
		\caption{Desired action refinement behavior}%
		\label{fig:desired refinement}%
	\end{subfigure}%
	\end{hideshowkeys}%
	\caption{Example for the (lack of) preservation of determinism in action refinement}
\end{figure}
At a lower level of abstraction, input $a$ is implemented by three consecutive inputs $1 \; 4 \; 1$, whereas input $b$ is implemented by action sequence $1 \; 4 \;  2$ (the ASCII encodings of $a$ and $b$ in octal format). An action refinement operator will replace the $a$-transition in Figure~\ref{ab} by a sequence of three consecutive transitions with labels $1$, $4$ and $1$, respectively, and will handle the $b$-transition in an analogous manner.  Thus, action refinement introduces a nondeterministic choice (\autoref{fig:existing refinement}), rather than the deterministic behavior that one would like to see (\autoref{fig:desired refinement}).
As a consequence of this and other limitations, refinement operators have not found much practical use \cite{GorrieriR01}.

Based on the observation that any action can be modeled as a state change, some authors (e.g. \cite{AL93,CGP99,Lam94}) prefer modeling formalisms in which the term ``action'' is only used informally, and Kripke structures rather than LTSs are used to model systems.
These state-based approaches have the advantage that a distinction between horizontal and vertical implementation relations is no longer needed, and a single implementation relation suffices.
Purely state-based approaches, however, are problematic in cases where we need to interact with a black-box system and (by definition) we have no clue about the state of this system. 
Black-box systems prominently occur in the areas of model-based testing \cite{Tret08} and model learning \cite{Vaa17}.  In these application areas, use of LTSs makes sense and there is a clear practical need for formalisms that allow engineers to relate actions at different levels of abstraction.

Van der Bijl et al \cite{BijlRT05}, for instance, observe that in model-based testing specifications are usually more abstract than the System Under Test (SUT). This means that generated test cases may not have the required level of detail, and often a single abstract action has to be translated (either manually or by an adaptor) to a sequence of concrete actions that are applied to the SUT. Van der Bijl et al \cite{BijlRT05} study a restricted type of action refinement in which a single input is refined into a sequence of inputs, and implement this in a testing tool.

Also in model learning, typically an adaptor is placed in between the SUT and the learner, to take care of the translation between abstract and concrete actions.
For example, in a case study on hand-held smartcard readers for Internet banking, Chalupar et al \cite{CPPR14} used abstract inputs that combine several concrete inputs in order to accelerate the learning process and reduce the size of the learned model. In particular, they introduced a single abstract input COMBINED\_PIN corresponding to a USB command, followed by a 4-digit PIN code, followed by an OK command.
Fiter\u{a}u-Bro\c{s}tean et al \cite{FiterauEtal20} used model learning for a comprehensive analysis of DTLS implementations, and found four serious security vulnerabilities, as well as several functional bugs and non-conformance issues.
Handshakes in (D)TLS are defined over flights of messages. Hence, (D)TLS entities are often expected to produce multiple messages
before expecting a response. During learning, Fiter\u{a}u-Bro\c{s}tean et al \cite{FiterauEtal20} used an adaptor that contracted  multiple messages from the SUT into a single abstract output.
Also in other case studies on TLS \cite{dRP15}, Wi-Fi \cite{StoneCR18} and SSH \cite{Verleg16,FiterauEtAl17}, multiple outputs from the SUT were contracted into a single abstract output.
Verleg \cite{Verleg16} used a single abstract input to execute the entire key re-exchange during learning higher layers of SSH. 

In this article, we provide answers to two fundamental questions:
(1) How can we formalize the concept of an \emph{adaptor} that translates between abstract and concrete actions?, and 
(2) Suppose the behavior of an SUT is described by an unknown, concrete model $\M$, and suppose a learner interacts with this SUT through an adaptor and learns an abstract model $\N$.  What can we say about the relation between $\M$ and $\N$?   

We answer the first question by introducing \emph{action codes}, a variation of the prefix codes known from coding theory \cite{berstel1985theory}. Action codes describe how high-level actions are converted into sequences of low-level actions, and vice versa. 
This makes them different from action refinements, which specify how high-level actions can be translated into low-level processes, but do not address the reverse translation.
Our notion of an action code captures adaptors that are used in practice, and in particular those described in the case studies listed above.  

In order to answer the second question we introduce, for each action code $\R$, a \emph{contraction} operator $\alpha_\R$ that turns a low-level model $\M$ into a high-level model by contracting concrete action sequences of $\M$ according to $\R$. 
We also introduce the left adjoint of $\alpha_\R$, the \emph{refinement} operator  $\rho_\R$ that turns a high-level model $\M$ into a low-level model by refining abstract actions of $\N$ according to $\R$. This refinement operator, for instance, maps the LTS of Figure~\ref{ab} to the LTS of \autoref{fig:desired refinement}.
We establish a Galois connection
	$\rho_\R(\N) \sqsubseteq \M ~ \Leftrightarrow ~ \N \sqsubseteq \alpha_\R(\M)$,
where $\sqsubseteq$ denotes the simulation preorder.
So if an abstract model $\N$ implements contraction $\alpha_\R(\M)$, then the refinement $\rho_\R(\N)$ implements concrete model $\M$, and vice versa.

In practice, we typically want to obtain an overapproximation of concrete model $\M$. To this end, we introduce the right adjoint of $\alpha_\R$, the \emph{concretization} operator $\gamma_\R$. This operator behaves like the refinement operator, but adds arbitrary behavior at intermediate points (cf. the demonic completion of \cite{BijlRT03}). We establish another Galois connection:
	$\alpha_\R(\M) \sqsubseteq \N$  $\Leftrightarrow$ $\M \sqsubseteq \gamma_\R(\N)$.
This connection is useful, because whenever we have established that $\alpha_\R(\M)$ implements (or conforms to) $\N$, it allows us to conclude that $\M$ implements (or conforms to) $\gamma_{\R} (\N)$.

We show that, in a setting of Mealy machines (Finite State Machines without finiteness requirement), an \emph{adaptor} can be constructed for any action code for which a winning strategy exists in a certain 2-player game. If a learner/tester interacts with an SUT via an adaptor generated from such an action code $\R$, and the SUT is modeled by Mealy machine $\M$, then from the learner/tester perspective, the composition of adaptor and SUT will behave like $\alpha_\R(\M)$.  Thus, if a learner succeeds to learn an abstract model $\N$ such that $\N \approx \alpha_\R(\M)$ then, using the Galois connections, the learner may conclude that $\rho_\R(\N) \sqsubseteq \M  \sqsubseteq \gamma_\R(\N)$.

The remainder of this article is structured as follows.
We start with a preliminary Section~\ref{sec: preliminaries} that introduces basic notations and results for LTSs.
Next, action codes and the contraction operator are introduced in Section~\ref{sec: action codes}.
After describing the refinement operator, we establish our first Galois connection
in Section~\ref{sec: refinements}.
Next we define concretization and establish our second Galois connection in Sections~\ref{sec: concretizations}. 
Section~\ref{sec: composition} explains how action codes can be composed, and shows that contraction and refinement commute with action code composition.
Section~\ref{sec: adaptors} describes how adaptors can be constructed from action codes.
Finally, Section~\ref{sec: discussion} contains a discussion of our results and identifies directions for future research.
Almost all proofs are formalized in Coq and can be accessed
\ifthenelse{\boolean{anonymous}}{%
anonymously via \url{https://}
}{%
via \url{https://arxiv.org/src/2301.00199/anc}
} -- we mark those results with a Coq icon {\coqicon}. Appendix~\ref{coqtoc} contains comments on the Coq formalization and
Appendix~\ref{appendix} contains full proofs (in natural language) and additional remarks.

\section{Preliminaries}
\label{sec: preliminaries}
If $\Sigma$ is a set of symbols then $\Sigma^*$ denotes the set of all finite words over $\Sigma$, and $\Sigma^+$ the set of all non-empty words.  We use $\epsilon$ to denote the empty word, so e.g.~$\Sigma^* = \Sigma^+\cup\set{\epsilon}$. Concatenation of words $u,w\in \Sigma^*$ is notated $u \cdot w$ (or simply $u\, w$). We write $u\le w$ if $u$ is a prefix of $w$, i.e.\ if there is $v\in \Sigma^*$ with $u\, v = w$. We write $| w |$ to denote the length of word $w$.

We use $f\colon X \partialto Y$ to denote a partial map $f$ from $X$ to $Y$ and
write $\dom(f) \subseteq X$ for its domain, i.e.~set of $x\in X$ on which $f$
is defined. The \emph{image} $\Im(f)$ of a partial map $f\colon X\partialto Y$ is the
set of elements of $Y$ it can reach: $\Im(f) := \set{f(x) \mid x\in \dom(f)} \subseteq Y$.

\begin{definition}[\coqref{LTS.v}{LTS}{}]
	For a set $A$ of action labels, a \emph{labeled transition system (LTS)} is a tuple
	$\M = \tuple{Q, q_0, \transto}$ where
	$Q$ is a set of states,
	$q_0 \in Q$ is a starting state, and
	$\mathord{\transto} \subseteq Q\times A\times Q$ is a transition relation.
	We write $\LTS(A)$ for the class of all LTSs with labels from $A$.
	We refer to the three components of an LTS $\M$ as $Q^\M$, $q_0^\M$ and $\transto[\M]$, respectively, and introduce the following notation:
\begin{itemize}
	\item $q\xtransto{a}q'$ denotes $(q,a,q')\in \mathord{\transto}$;
	$q\xtransto{a}$ denotes that there is some $q'$ with $q\xtransto{a}q'$;
	\item $q\xTransto{w}q'$ for $w\in A^*$ denotes that there are finite sequences
		$a_1,\ldots,a_n\in A$, $r_0,\ldots,r_n\in Q$ such that $w=a_1\cdots a_n$, 
		and $r_0 = q$, $r_n=q'$ and $r_{i-1}\xtransto{a_i} r_i$ for all $1\le i\le n$;
	\item $q\xTransto{w}$ denotes that there is $q'$ such that $q\xTransto{w}q'$;
	\item $q \in Q$ is \emph{reachable} if there is $w \in A^*$ such that $q_0 \xTransto{w}q$.
	\end{itemize}
\end{definition}
A special class of LTSs that is frequently used in conformance testing and model learning are \emph{Mealy machines}.  Mealy machines with a finite number of states are commonly referred to as \emph{Finite State Machines}.

\tikzset{
	mealy square example/.style={
		unreachable/.style={
			opacity=0.3,
		},
		q0/.style={},
		q1/.style={},
		q2/.style={},
		q3/.style={},
		square transitions/.style={},
	}
}%
\newcommand{\mealySquareExample}{%
	\begin{scope}[mealy square example]
	\node[state, initial,q0] (q0) {$q_0$};
	\node[state, right of=q0,q1] (q1) {$q_1$};
	\node[state, below of=q1,q2] (q2) {$q_2$};
	\node[state, below of=q0,q3] (q3) {$q_3$};
	\begin{scope}[transparency group, square transitions]
	\draw[transition, bend angle=15]
	(q0) edge[bend right] node[swap]{$b/0$} (q1)
	(q0) edge[bend right] node[swap]{$a/0$} (q3)
	(q1) edge[bend right] node[swap]{$a/0\;$} (q2)
	(q1) edge[bend right] node[swap]{$b/0$} (q0)
	(q2) edge[bend right] node[swap]{$b/0$} (q1)
	(q2) edge[bend right] node[swap]{$a/0$} (q3)
	(q3) edge[bend right] node[swap]{$a/0$} (q2)
	(q3) edge[bend right] node[swap]{$b/1$} (q0);
	\end{scope}
	\end{scope}
}%
\begin{iflong}%
\pagebreak%
\FloatBarrier%
\end{iflong}%
\begin{wrapfigure}[8]{r}[0mm]{.3\textwidth}%
	\ifthenelse{\boolean{longversion}}{
	}{%
		\vspace{-5mm}%
	}%
	\hspace*{0pt}\hfill%
	\begin{tikzpicture}[lts,node distance=2.3cm]
	\mealySquareExample
	\end{tikzpicture}%
	\vspace{-1mm}%
	\caption{A Mealy machine.\ifthenelse{\boolean{longversion}}{
		\hspace*{-2pt}%
	}{}}%
	\label{fig:example_mealy_machine}%
\end{wrapfigure}%
\hspace*{0pt}\vspace{-\baselineskip}%
\begin{definition}
  For non-empty sets of inputs $I$ and outputs $O$, a (non-deterministic)
  \emph{Mealy machine} $\M\in \LTS(I\times O)$ is an LTS where the
  labels are pairs of an input and an output.
  We write $q \xtransto{i / o} q'$ to denote that $(q,(i,o),q')\in \transto$.
Whenever we omit a symbol in predicate $q \xrightarrow{i / o} q'$ this is quantified existentially.  Thus,
$ \xtransto{i/o}$ if there are $q$ and $q'$ s.t.\  $q \xtransto{i / o} q'$,
$q \xtransto{i/} q'$ if there is an $o$ s.t.\  $q \xtransto{i / o} q'$, and
$q \xtransto{i/}$ if there is a $q'$ s.t.\  $q \xtransto{i/} q'$.
\end{definition}

\begin{example}[\coqref{MealyExample.v}{ExampleMealy}{}]
Figure \ref{fig:example_mealy_machine} visualizes a simple Mealy machine with inputs $\{a, b\}$ and outputs $\{0, 1\}$.
The machine always outputs $0$ in response to an input, except in one specific situation. Output $1$ is produced in response to input $b$ if
the previous input was $a$ and the number of preceding inputs is odd.
The machine has four states $q_0, q_1, q_2$ and $q_3$, with starting state $q_0$ marked by an incoming arrow.
In states $q_0$ and $q_2$ the number of preceding inputs is always even, whereas in states $q_1$ and $q_3$ it is always odd.
In states $q_2$ and $q_3$ the previous input is always $a$, whereas in states $q_0$ and $q_1$ either the previous input is $b$, or no input has occurred yet.
Thus, only in state $q_3$ input $b$ triggers output $1$.
\end{example}

We introduce some notation and terminology for LTSs.

\begin{definition}[\coqref{LTS.v}{deterministic}{In Coq, we have the notion of determinism parametric in a relation $\SI\subseteq A\times A$ such that we can uniformly treat determinism in the sense of LTSs and determinism in the sense of Mealy machines (determinism in the \emph{input}-component of the action labels). This relation will be described in \autoref{sec: concretizations}. \\ The other notions (tree-shaped, leaf, grounded) are formalized in \UScore{TreeShapedCode.v}.}]
Let $\M = \tuple{Q,q_0,\transto} \in \LTS(A)$ be an LTS. 
We say that
\begin{itemize}
\item
$\M$ is \emph{deterministic} if, whenever $q \xtransto{a}$ for some $q$ and $a$, there is a unique $q'$ with $q \xtransto{a} q'$.
\item
$\M$ is a \emph{tree-shaped} if each state $q \in Q$ can be reached via a unique sequence of transitions from state $q_0$.

\item $q\in Q$ is a \emph{leaf}, notated $q\notransto$, if there is no $a\in A$ with $q\xtransto{a}$.
\item $\M$ is \emph{grounded} if every state $q\in Q$ has a path to a leaf.
\end{itemize}
\end{definition}

We can now define the set of traces of an LTS:
\begin{definition}[\coqref{LTS.v}{traces}{}]
	Let $\M = \tuple{Q,q_0,\transto} \in \LTS(A)$.
	A word $w\in A^*$ is a \emph{trace} of state $q \in Q$ if $q \xTransto{w}$, and a \emph{trace} of $\M$ is it is a trace of $q_0$. We write $\obs(\M)$ for the set $\set{w\in A^*\mid q_0\xTransto{w}}$ of all traces of $\M$.
\end{definition}

\begin{definition}[Simulation, \coqref{LTS.v}{Simulation}{}]
	For $\M, \N \in \LTS(A)$,
	a \emph{simulation} from $\M$ to $\N$ is a relation $S \subseteq Q^{\M} \times Q^{\N}$ such that
	\begin{enumerate}
		\item 
		$q_{0}^{\M} \mathrel{S} q_{0}^{\N}$ and
		\item 
		if $q_1 \mathrel{S} q_2$ and $q_1 \xtransto[\M]{a} q'_1$ then there exists a state $q'_2$ such that $q_2 \xtransto[\N]{a} q'_2$ and $q'_1 \mathrel{S} q'_2$.
	\end{enumerate}
We write $\M \sqsubseteq \N$ if there exists a simulation from $\M$ to $\N$.
\end{definition}

It is a classical result that trace inclusion coincides with the simulation preorder for deterministic labeled transition systems (see e.g. \cite{LV95}):

\begin{lemma}[\coqref{LTS.v}{simulation_iff_trace_inclusion_for_deterministic}{}]
	\label{La simulation}
	For all $\M, \N \in \LTS(A)$ where $\N$ is deterministic: 
	 $\obs(\M) \subseteq \obs(\N)$ iff $\M \sqsubseteq \N$.
\end{lemma}

We will often consider LTSs up to isomorphism of their reachable parts:

\begin{definition}[Isomorphism, \coqref{LTS.v}{Isomorphism}{%
	In Coq, we implicitly restrict to reachable states by only requiring $f$ to
	be a left- and right-unique relation $f\subseteq Q^\M\times Q^\N$, but without
	any requirement that $f$ is total. Together with the other properties of $f$,
	this relation restricts to an isomorphism between the reachable states.%
}]
	\label{def:isomorphism}
	For $\M, \N \in \LTS(A)$,
	an \emph{isomorphism} from $\M$ to $\N$ is a bijection $f \colon Q_R^\M \rightarrow Q_R^\N$, where:
	\begin{enumerate}
	\item $Q_R^\M \subseteq Q^\M$ and $Q_R^\N\subseteq Q^\N$ are the subsets of reachable states in $\M$ and $\N$, respectively;
	\item $f(q_0^\M) = q_0^\N$, and
	\item $q \xtransto[\M]{a} q'$ iff $f(q) \xtransto[\N]{a} f(q')$, for all $q, q' \in Q_R^\M$, $a \in A$.
	\end{enumerate}
	We write $\M \cong \N$ if there exists an isomorphism from $\M$ to $\N$.
\end{definition}
Note that $\cong$ is an equivalence relation on $\LTS(A)$, and that $\M \cong \N$ implies $\M \sqsubseteq \N$, since each isomorphism (when viewed as a relation) is trivially a simulation.

\section{Action Codes}
\label{sec: action codes}
Action codes describe how to translate between two action label alphabets, for example from $A$ to $B$. Intuitively, we understand the first alphabet $A$ as the actions at the lower, concrete level, and the second alphabet $B$ as the actions at the higher, more abstract level. In an action code, a single abstract action $b\in B$ corresponds to a finite, non-empty sequence of concrete actions $a_1\cdots a_n$ in $A$.
Essentially, action codes are just a special type of \emph{prefix codes}~\cite{berstel1985theory}.
We provide two equivalent definitions of action codes: one via tree-shaped LTSs and one via partial maps.

\begin{definition}[Action code, \coqref{TreeShapedCode.v}{TreeShapedCode}{}]
	\label{def:actioncode}
	For sets of action labels $A$ and $B$, a \emph{(tree-shaped) action code} $\R$ from $A$ to $B$
	is a structure $\R = \tuple{\M, l}$, with $\M = \tuple{R,r_0,\transto} \in
	\LTS(A)$ a deterministic, tree-shaped LTS with $L$ being the set of
	non-root leaves $L \subseteq R \setminus \set{ r_0 }$ and an injective
	function $l\colon L \to B$. We write $\Code(A,B)$ for all action codes from $A$ to $B$.
\end{definition}
The injectivity of $l$ and the tree-shape ensure that every abstract
$b\in B$ is represented by at most one $w\in A^+$.

\begin{example}
	Figure~\ref{ascii} shows an action code for a fragment of the ASCII encoding in octal format, e.g., $1 \, 1 \, 5$ encodes the letter \ascii{M}, $1 \, 4 \, 5$ encodes the letter \ascii{e}, etc.
\end{example}
	\begin{figure}[th]
		\begin{minipage}[b]{.28\textwidth}%
			\begin{tikzpicture}[lts,empty states, node distance=18cm,
			                    x=4mm,y=15mm]
			\node[initial above, state] (1) at (0,-0.2) {\hidestatename{$r_0$}};
			\node[state] (2) at (0,-1) {\hidestatename{$r_1$}};
			\node[state] (3) at (-3,  -2.3) {\hidestatename{$r_2$}};
			\node[state] (4) at (-0.9,-2.3) {\hidestatename{$r_3$}};
			\node[state] (5) at ( 0.9,-2.3) {\hidestatename{$r_4$}};
			\node[state] (6) at ( 3,  -2.3) {\hidestatename{$r_5$}};
			\begin{scope}[
				every label/.append style={
					text height=3mm,
				},
				node distance=14mm,
				]
			\node[state, label={below:$\ascii M$}] (7) [xshift=-5mm,below of=3] {\hidestatename{$r_6$}};
			\node[state, label={below:$\ascii e$}] (8) [xshift=-5mm,below of=4] {\hidestatename{$r_7$}};
			\node[state, label={below:$\ascii a$}] (9) [xshift=5mm,below of=4] {\hidestatename{$r_8$}};
			\node[state, label={below:$\ascii l$}] (10) [xshift=5mm,below of=5] {\hidestatename{$r_9$}};
			\node[state, label={below:$\ascii y$}] (11) [xshift=5mm,below of=6] {\hidestatename{$r_{10}$}};
			\end{scope}
			
			\path[transition,auto]
			(1) edge node {$1$} (2)
			(2) edge node[swap] {$1$} (3)
				edge node[pos=0.8,swap] {$4$} (4)
				edge node[pos=0.8] {$5$} (5)
				edge node {$7$} (6)
			(3) edge node[swap] {$5$} (7)
			(4) edge node[swap] {$5$} (8)
				edge node {$1$} (9)
			(5) edge node {$4$} (10)
			(6) edge node {$1$} (11)
			;
			\end{tikzpicture}\\
			~%
		\end{minipage}%
		\hfill%
		\begin{minipage}[b]{.67\textwidth}%
		\newcommand{\tikzSubtreeOnReady}[1]{%
			child[sibling distance=10mm,level distance=13mm] {
				node[label=below:{\imgcoffee/#1}]{}
				edge from parent node[swap,pos=0.7] {\imgcoffee/\\ coffee}}
			child[sibling distance=10mm,level distance=13mm] {
				node[label=below:{\imgespresso/#1}]{}
				edge from parent node[pos=0.7] {\imgespresso/\\ espresso}}
		}
		\begin{tikzpicture}[lts,empty states]
			\node[initial above,state] (root) {};
			\begin{scope}[
				every node/.style={
					state,
				},
				edge from parent/.style={
					transition,
					every node/.style={
						align=center,
						font=\footnotesize\itshape,
						inner sep=0pt,
						outer sep=0pt,
					},
				},
				slola/.style={
					sloped,above,
					inner ysep=1pt,
				},
				sibling distance=15mm,
				level distance=11mm,
				level 1/.style = {sibling distance = 4cm},
				level 2/.style = {sibling distance = 25mm},
				level 3/.style = {level distance=13mm,sibling distance = 18mm},
			]
			\node (r) at (root.center) {}
				child {node(switch on ready) {}
					\tikzSubtreeOnReady{0}
					edge from parent node[swap] {switch\_on/\\ ready}
				}
				child {node{}
					child { node(add water ready) at ([yshift=-36mm,xshift=4mm]switch on ready) {}
						\tikzSubtreeOnReady{1}
						edge from parent node[slola] {add\_water/ready}
					}
					child { node[xshift=5mm] {}
						child { node[xshift=-15mm,yshift=-7mm] (add beans ready) {}
							\tikzSubtreeOnReady{2}
							edge from parent node[swap] {add\_beans/\\ ready}}
						child { node{}
							child {node{}
								\tikzSubtreeOnReady{3}
								edge from parent node[swap] {remove\_waste/~~\\ready}}
							edge from parent node {add\_beans/\\tray\_full}}
						edge from parent node {add\_water/\\ need\_beans}
					}
					edge from parent node {switch\_on/\\ need\_water}
				}
			;
			\end{scope}
		\end{tikzpicture}
		\end{minipage}%
		\\%
		\begin{minipage}{.5\textwidth}%
		\caption{Action code for a fragment of ASCII.}
		\label{ascii}
		\end{minipage}%
		\hfill%
		\begin{minipage}{.45\textwidth}%
		\caption{Action code for a coffee machine.}
		\label{D4}
		\end{minipage}%
	\end{figure}

\begin{example}
	Figure~\ref{D4} shows an action code for the activity of getting a cup of coffee or espresso, in the special case of Mealy machines, i.e.~where $A = I\times O$ and $B=I'\times O'$ are sets of input/output-pairs.  Rather than the full sequence of interventions that is required in order to get a drink, the abstract input/output pair only reports on the type of drink that was ordered and the number of interventions that occurred.
	\takeout{} %
\end{example}
The definition of action codes as LTSs allows
an intuitive visualization. For easier mathematical reasoning, we characterize
action codes also in terms of maps:
\begin{definition}[\coqref{CodeMap.v}{Code}{%
	In the formlization, the type of a code map is $f\colon B\partialto A^*$
	(instead of $B\partialto A^+$)
	and then in addition to prefix-freeness, we require this map never returns
	the empty word. The reason for this decision is that $A^*$  is just lists over $A$ -- a standard data type with good standard library support.}]
	A \emph{(map-based) action code} from $A$ to $B$ is a partial map
	$f\colon B\partialto A^+$ which
	is \emph{prefix-free}, by which we mean that for all $b,b'\in \dom(f)$,
		\begin{equation}
			f(b) \le f(b') \qquad\text{implies}\qquad b=b'.
			\label{prefix-free}
		\end{equation}
\end{definition}

In the following, we show that these prefix-free partial maps bijectively
correspond to the tree-shaped LTSs:

\begin{lemma}[\coqref{TreeShapedCode.v}{EveryTreeShapedCodeInducesSomeMapBasedCode}{Being in a constructive setting, we have an explicit assumption that decides for each $b\in B$, whether there exists a leaf with label $b$. Also, the uniqueness of the map-based code
is stated in a separate Lemma \UScore{TreeDefinesMapUniquely}, to avoid the use of the functional extensionality axiom.}]
	\label{treeCodeToMap}
	Every tree-shaped action code $\R\in \Code(A,B)$ induces a unique map-based
	action code $f\colon B\partialto A^+$ with the property that for all $b\in B, w\in A^+$:
	\begin{equation}
		f(b) = w
		\qquad\text{ iff }\qquad
		\exists r \in L\colon
		r_0\xTransto[\R]{w} r,~~ l(r) = b
		\label{defMapR}
  \end{equation}
\end{lemma}

\begin{lemma}[\coqref{TreeShapedCode.v}{MapToTree}{%
In the definition of the LTS, we have to be careful such that 1.~Coq allows
us to extract witnesses and 2.~we still don't have duplicate states.
The assumption of groundedness is only necessarily in the uniqueness (and not in existence), which is proven as a separate Lemma \UScore{MapDefinesTreeUniquely}.}]
	\label{mapCodeToTree}
	For each map-based action code $f\colon B\partialto A^+$, there is (up to isomorphism) a unique
	tree-shaped action code $\R\in \Code(A,B)$ which is grounded and satisfies \eqref{defMapR}.
\end{lemma}

\begin{example}[\coqref{TreeShapedCode.v}{groundedness_is_needed_for_uniqueness}{}]
For the  uniqueness in \autoref{mapCodeToTree}, we use groundedness,
because for $A= \set{a}$ and any $B$, the action codes
\[
	\R :=
	\big(
	\begin{tikzpicture}[lts,x=1.5cm,baseline={([yshift=-3pt]r0.base)}, empty states]
		\node[initial,state,alias=lastnode] (r0) {\hidestatename{$r_0$}};
		\foreach \index in {1,2} {
			\node[state] (r\index) at (\index,0) {\hidestatename{$r_{\index}$}};
			\path[transition] (lastnode) -- node {$a$} (r\index);
			\coordinate (lastnode) at (r\index.east);
		};
		\node (dots) at ([xshift=14mm]lastnode) {$\cdots$};
		\path[transition] (lastnode) -- node {$a$} (dots);
	\end{tikzpicture}
	\big)
	\quad\text{and}\quad
	\S :=
	\big(
	\begin{tikzpicture}[lts,x=2cm,baseline={([yshift=-3pt]r0.base)},empty states]
		\node[initial,state,alias=lastnode] (r0) {\hidestatename{$q_0$}};
	\end{tikzpicture}
	\big).
\]
both have no non-root leaves, and so they both
induce the empty partial map $f\colon B\partialto A^+$ via \autoref{treeCodeToMap}.%
\begin{iflong}
This~$f$ is undefined for all $b\in B$.
And indeed, $\R$ and $\S$ are not isomorphic. The issue is that while
the finite $\S$ is grounded, the infinite $\R$ is not grounded. So $\R$
contains subtrees which do not contribute anything to the partial map $f$
but which hinder the existence of an isomorphism.
\end{iflong}
\end{example}

Having shown the correspondence between tree-shaped and map-based action codes $\Code(A,B)$,
we can switch between the two views in proofs. Mostly, we use the tree-shaped
version for visualization and the map-based version for mathematical reasoning.

Consider a concrete $\M \in \LTS(A)$, together with an action code $\R$ from $A$ to $B$.
We can construct an abstract LTS for the action labels $B$ by walking through $\M$ with seven-league boots, repeatedly choosing input sequences that correspond to runs to some leaf of $\R$, and then contracting this sequence to a single abstract transition.

\begin{notation}
	In the rest of the paper, we introduce operators $\alpha_\R$, $\rho_\R$,
	$\gamma_\R$ on LTSs, involving an action code $\R$. Whenever the action code
	$\R$ is clear from the context, we omit the index and simply speak of
	operators $\alpha$, $\rho$, $\gamma$ for the sake of brevity.
\end{notation}

\begin{definition}[Contraction, \coqref{Contraction.v}{contraction}{In the Coq version,
	we do not restrict to reachable states, because it makes the reasoning easier
	to if $\M$ and $\alpha_\R(\M)$ have the same states. For the visualization of
	the examples, it is easier to omit unreachable states. In our notion of
	isomorphism (\autoref{def:isomorphism}), every LTS is isomorphic to its restriction
	to reachable states.}]
	\label{Def contraction}
For each action code $\R \in \Code(A,B)$, the \emph{contraction} operator $\alpha_\R\colon \LTS(A) \to \LTS(B)$ is defined as follows.	
For $\M \in \LTS(A)$, the LTS $\alpha_\R(\M)$ has states
$Q^{\alpha(\M)}\subseteq Q^\M$ and transitions $\transto[\alpha(\M)]$ defined inductively by the rules \eqref{Rule A1} and \eqref{Rule A2},
for all $q, q' \in Q^\M$, $b \in B$.%
\makeatletter
\[
	\infer[\rulelabel{Rule A1}{$1_\alpha$}]
		{q_0^\M \in Q^{\alpha(\M)}}{}
	\qquad\qquad
	\infer[\rulelabel{Rule A2}{$2_\alpha$}]
		{q \xtransto[\alpha(\M)]{b} q',  \quad\quad q' \in Q^{\alpha(\M)}}
		{q \in Q^{\alpha(\M)}, & b\in \dom(\R), & q \smash{\xTransto[\M]{\R(b)}}\, q'}
\]
\makeatother
The initial state $q_0^{\alpha(\M)} := q_0^\M$ is the same as in $\M$.
\end{definition}

\begin{example}
		Figures~\ref{example contraction both} shows two examples of action codes and the contractions obtained when we apply them to the Mealy machine of Figure~\ref{fig:example_mealy_machine} (with the original machine shaded in the background).
		The examples illustrate that by choosing different codes we may obtain completely different abstractions of the same LTS.
		\begin{figure}[t]
		\begin{hideshowkeys}
		\def\codeExamplesSiblingOffset{6mm}
		\begin{subfigure}[t]{.48\textwidth}%
			\begin{tikzpicture}[
				lts, empty states,node distance=1.3cm,
				baseline=(current bounding box.center),
			]
			\node[initial above, state] (1){\hidestatename{$r_0$}};
			\node[state] (2) [below of=1,xshift=-\codeExamplesSiblingOffset] {\hidestatename{$r_1$}};
			\node[state] (3) [below of=1,xshift=\codeExamplesSiblingOffset] {\hidestatename{$r_2$}};
			\node[state, label={below:$A/\mathbf{0}$}] (4) [below of=2] {\hidestatename{$r_3$}};
			\node[state, label={below:$B/\mathbf{0}$}] (5) [below of=3] {\hidestatename{$r_4$}};
			
			\path[transition]
			(1) edge node[swap] {$a/0$} (2)
				edge node {$b/0$} (3)
			(2) edge node[swap] {$a/0$} (4)
			(3) edge node {$b/0$} (5)
			;
			\end{tikzpicture}
			\hfill%
			\begin{tikzpicture}[
				lts,
				baseline=($ (q0.center) !.5! (q2.center) $),
				mealy square example/.append style={
					q1/.append style = {unreachable},
					q3/.append style = {unreachable},
					square transitions/.append style = {
						unreachable,
						every node/.append style={
							overlay,
							opacity=0,
						},
					},
				},
			]
			\mealySquareExample
			\draw[
				transition,
				bend angle=15,
				every node/.append style={
					white background node,
				},
			]
				(q2) edge[white glow,loop, looseness=10,out=165,in=195] node[below,yshift=-2mm] {$A/\mathbf{0}$} (q2)
				(q0) edge[white glow,bend right] node[sloped,below] {$A/\mathbf{0}$} (q2)
				(q0) edge[white glow,loop, looseness=10,out=-15,in=15] node[above,yshift=2mm]{$B/\mathbf{0}$} (q0)
				(q2) edge[white glow,bend right] node[sloped,above]{$B/\mathbf{0}$} (q0)
			;
			\draw[draw=none] ([yshift=-2mm]q3.south) circle (1pt);
			\end{tikzpicture}
		\end{subfigure}%
		\hfill%
		\begin{subfigure}[t]{.48\textwidth}%
			\begin{tikzpicture}[
				lts,empty states,node distance=1.2cm,
				baseline=(current bounding box.center),
				]
			\node[initial above, state] (1){\hidestatename{$r_0$}};
			\node[state] (2) [below of=1,xshift=-\codeExamplesSiblingOffset] {\hidestatename{$r_1$}};
			\begin{scope}[every label/.append style={xshift=1mm,overlay}]
			\node[state, label={below:$B/\mathbf{0}$}] (3) [below of=1,xshift=\codeExamplesSiblingOffset] {\hidestatename{$r_2$}};
			\end{scope}
			\node[state, label={below:$C/\mathbf{0}$}] (4) [below of=2,xshift=-\codeExamplesSiblingOffset] {\hidestatename{$r_3$}};
			\node[state, label={below:$C/\mathbf{1}$}] (5) [below of=2,xshift=\codeExamplesSiblingOffset] {\hidestatename{$r_4$}};
			
			\path[transition]
			(1) edge node[swap] {$a/0$} (2)
				edge node {$b/0$} (3)
			(2) edge node[swap] {$b/0$} (4)
			 	edge node {$b/1$} (5)
			;
			\end{tikzpicture}
			\hfill%
			\begin{tikzpicture}[
				lts,
				baseline=($ (q0.center) !.5! (q2.center) $),
				mealy square example/.append style={
					q2/.append style = {unreachable},
					q3/.append style = {unreachable},
					square transitions/.append style = {
						unreachable,
						every node/.append style={
							overlay,
							opacity=0,
						},
					},
				},
			]
			\mealySquareExample
			\draw[transition,bend angle=15]
			(q0) edge[white glow,bend right] node[swap] {$B/\mathbf{0}$} (q1)
			(q0) edge[white glow,loop,looseness=10,out=270-15,in=270+15]
			           node[right,outer sep=3pt,pos=0.7]{$C/\mathbf{1}$} (q0)
			(q1) edge[white glow,bend right] node[swap] {$B/\mathbf{0}$} (q0)
			(q1) edge[white glow,loop,looseness=10,out=270-15,in=270+15]
			           node[left,outer sep=3pt,pos=0.3]{$C/\mathbf{0}$} (q1)
			;
			\end{tikzpicture}
		\end{subfigure}%
		\\
		\begin{subfigure}[t]{.48\textwidth}%
			\caption{An action code $\R$ together with $\alpha_\R(\M)$}
			\label{example contraction}
		\end{subfigure}%
		\hfill%
		\begin{subfigure}[t]{.48\textwidth}%
			\caption{Another code $\S$ together with $\alpha_\S(\M)$}
			\label{example contraction2}
		\end{subfigure}%
		\end{hideshowkeys}
		\caption{The resulting contraction of
		the LTS $\M$ from \autoref{fig:example_mealy_machine} for different action codes.}
		\label{example contraction both}
	\end{figure}
\end{example}

The next proposition asserts that we can view $\alpha_\R$ as a monotone function $\alpha_\R\colon \LTS(A)\to \LTS(B)$ between preordered classes.	

\begin{proposition}[Monotonicity, \coqref{LTS.v}{contraction_monotone}{}]
	\label{contraction monotone}
	\twnote{}
	For every action code $\R\in \Code (A,B)$, whenever $\M \sqsubseteq \N$ for $\M,\N \in \LTS(A)$, then $\alpha_\R (\M) \sqsubseteq \alpha_\R (\N)$ in $\LTS(B)$.
\end{proposition}

\section{Refinements}
\label{sec: refinements}
Now that we have introduced the contraction $\alpha_\R$ of an LTS for a code $\R$, it is natural to consider an operation in the other direction, which we call the \emph{refinement} $\rho_\R$. 
Intuitively, refinement replaces each abstract transition $q \xtransto{b} q'$ by a sequence of concrete transitions, as prescribed by $\R$.

\begin{definition}[Refinement, \coqref{Refinement.v}{refinement}{}]
	\label{def: refinement}
	For each action code $\R\in \Code(A, B)$, we define the \emph{refinement} operator $\rho_\R\colon \LTS(B) \to \LTS(A)$ as follows.
	For $\M\in \LTS(B)$, the LTS $\rho_\R(\M) \in \LTS(A)$ has a set of states
	\[
		Q^{\rho(\M)}
		:=
		\set{
			(q,w)
			\in Q^\M\times A^*
			\mid
			w=\epsilon
			\text{ or }
			(\text{there is } b \text{ with } q\xtransto[\M]{b}\text{ and }
			w \lneqq \R(b))
		}
	\]
	and the initial state $(q_0^\M,\epsilon)$.
	The transition relation $\transto[\rho(\M)]$ is defined by the following rules:
	\[
		\infer[\rulelabel{rho mid}{$1_\rho$}]
			{(q,w) \xtransto[\rho(\M)]{a} (q, wa)}{(q,wa)\in Q^{\rho(\M)}}
		\qquad\qquad
		\infer[\rulelabel{rho end}{$2_\rho$}]
			{(q,w)\xtransto[\rho(\M)]{a} (q',\epsilon)}
			{q \xtransto[\M]{b} q' & wa=\R(b)}
	\]
\end{definition}
Intuitively, whenever $\rho(\M)$ is in state $(q,w)$, then this corresponds to being in state $q$ in
the abstract automaton $\M\in \LTS(B)$ and having observed the actions $w\in A^*$
so far. However, we have insufficiently many actions for finding an abstract
transition $q\xtransto[\M]{b} q'$ with $w = \R(b)$ because $w$ is still to
short. Nevertheless, whenever $\rho(\M)$ admits a transition to a state $(q,w)$
with $w\neq \epsilon$, then we know that we can eventually complete $w$ to a
sequence corresponding to an abstract transition: there exist
at least one $q\xtransto[\M]{b} q'$ for some $b\in \dom(\R)$ with $w \le \R(b)$.
If the abstract system $\M$ is non-deterministic, then there may be multiple abstract transitions that match
in the final rule \eqref{rho end}, but the transitions produced by rule \eqref{rho mid} are deterministic.

\begin{example}
	Figure~\ref{example refinement} shows an example application of a refinement operator that replaces the actions of the LTS $\M$ on the left by their ASCII encoding in octal format, as prescribed by the action code from Figure~\ref{ascii}.
	The initial state is $(q_0,\epsilon)$, corresponding to $q_0$ in $\M$.
	Since $\M$ contains abstract labels $\ascii{M}$ and $\ascii{a}$, with $\R(\ascii M) = 1 \, 1 \, 5$ and $\R(\ascii a) = 1 \, 4 \, 1$, we need to introduce additional states for having read $1$, $1 \, 1$, and $1 \, 4$, because those are the sequences of $A$-actions before we have observed a
	sequence $\R(b)\in A^+$ for some $b\in B$.
	\begin{figure}[t]\centering%
	\begin{minipage}[b]{.55\textwidth}%
		\begin{tikzpicture}[lts,node distance=2cm,baseline=(1.base)]
		\node[initial, state] (1){$q_0$};
	
		\path[transition,auto,overlay]
		(1) edge[loop,out=10,in=70,looseness=6] node[swap] {\ascii{M}} (1)
			edge[loop,out=-10,in=-70,looseness=6] node {\ascii{a}} (1)
		;
		\end{tikzpicture}%
		\hfill%
		\begin{tikzpicture}[lts,x=14mm,y=6mm,baseline=(r0.base)]
		\node[state, initial] (r0) at (-1.7,0) {$q_0, \epsilon$};
		\node[state] (r1) at (-0.3,0) {$q_0, 1$};
		\node[state] (r2) at (1,1) {$q_0, 11$};
		\node[state] (r3) at (1,-1) {$q_0, 14$};
		\draw[transition,bend angle=19,overlay]
		(r0) edge node{$1$} (r1)
		(r1) edge node{$1$} (r2)
		(r1) edge node[swap]{$4$} (r3)
		(r2) edge[bend right] node[swap] {$5$} (r0)
		(r3) edge[bend left] node{$1$} (r0)
		;
		\end{tikzpicture}%
	\caption{LTS and its refinement w.r.t~$\R$ of Figure~\ref{ascii}.}
	\label{example refinement}
	\end{minipage}%
	\hfill%
	\begin{minipage}[b]{.41\textwidth}\centering
		\begin{tikzcd}[column sep=14mm]
			\rho_\R(\N) \sqsubseteq \M
			\arrow[Rightarrow,shorten <=3pt,shift left=2]{r}[yshift=2mm,font=\upshape\scriptsize]{\text{If }\N\in\LTS(\dom(\R))}
			& \N \sqsubseteq \alpha_\R(\M)
			\arrow[Rightarrow,shorten <=3pt,shift left=2]{l}[yshift=-2mm,font=\upshape\scriptsize]{\text{If }\M\text{ is deterministic}}
		\end{tikzcd}
		\caption{\autoref{galois_connection refinement}}
	\end{minipage}
\end{figure}%
\end{example}%
A more visual explanation of $\rho_\R(\M)$ is the following: for every state $q\in Q^\M$, we consider the outgoing transitions $\set{q\xtransto[\M]{b} q'\mid b\in B, q'\in Q^\M}$ and labels $B'\subseteq B$ that appear in it.
Then, this outgoing-transition structure is replaced with
(a copy of) the minimal subgraph of the tree $\R$ containing all leaves with labels in $B'$.

Like contraction, the refinement operation also preserves the simulation preorder.

\begin{proposition}[Monotonicity, \coqref{Refinement.v}{refinement_monotone}{}] \label{rhoMonotone}
  For all action codes $\R\in \Code(A,B)$, if $\M\sqsubseteq \N$ in $\LTS(B)$, then
  $\rho_\R(\M) \sqsubseteq \rho_\R(\N)$ in $\LTS(A)$.
\end{proposition}

As $\R$ is deterministic, applying $\rho_\R$ on a deterministic LTS results in a deterministic LTS:

\begin{proposition}[Refinement preserves determinism, \coqref{Refinement.v}{refinement_preserves_determinism}{}]
	\label{rho deterministic}
	 For every action code $\R\in \Code(A,B)$,
	if $\M\in \LTS(B)$ is deterministic, then $\rho_\R(\M)\in \LTS(A)$ is
	deterministic, too.
\end{proposition}

\begin{theorem}[Galois connection,
\coqref{Refinement.v}{refinement_galois_forward / refinement_galois_backward}{In Coq, the assumption of the first direction is spelled out as: $\R$ is defined for all action codes that appear in $\N$}]\label{galois_connection refinement}
	For $\R\in \Code(A, B)$, $\N\in \LTS(B)$, and $\M\in \LTS(A)$:
	\begin{enumerate}
	\item If $\N$ is in the subclass $\LTS(\dom(\R))\subseteq \LTS(B)$, then $\rho_\R(\N) \sqsubseteq \M$ implies $\N \sqsubseteq \alpha_\R(\M)$.
	\item If $\M$ is deterministic, then $\N \sqsubseteq \alpha_\R(\M)$ implies $\rho_\R(\N) \sqsubseteq \M$.
	\end{enumerate}
\end{theorem}
\begin{iflong}
The condition in the first direction means that $\N\in\LTS(B)$ only makes use of action labels in the subset $\dom(\R)\subseteq B$. Hence, in the proof, we can consider
$\R$ to be a total map $\dom(\R)\to A^+$.
\end{iflong}
\begin{remark}
	If we wanted to support non-deterministic $\M$, we can consider a less-pleasant $\rho'_\R$ that
	replaces every $q\xtransto{b}q'$ for $\R(a_1\cdots a_n) = b$ with literally a
	sequence $q\xtransto{a_1}\cdots\xtransto{a_n} q'$.  Thus, $\rho_\R'$ would
	rather create a system as in \autoref{fig:existing refinement} whereas $\rho_\R$
	creates a system as in \autoref{fig:desired refinement}.
	However, such an operator $\rho_\R'$ does not preserve determinism.
\end{remark}
\begin{iflong}
\begin{remark} \label{why functional}
	In the proof of the Galois connection, we make use of the fact that our
	action codes are functional, i.e.~that every $b\in B$ is encoded by at most
	one $w\in A^*$. We would allow multiple, then
	one can show that $\alpha$ can not have a left-adjoint (details in appendix).
\end{remark}
\end{iflong}

\section{Concretizations}
\label{sec: concretizations}
In this section, we consider another method of transforming an abstract system into a concrete one: the \emph{concretization} operator. Whereas refinement is the left adjoint of contraction (Theorem~\ref{galois_connection refinement}), this section will establish that concretization is the right adjoint (Theorem~\ref{gamma galois}) of contraction.
Whereas for refinement we omitted transitions for which the action code $\R$ was not defined,
for concretization we add transitions to a new \emph{chaos} state \cite{Ho85} in which any action may occur. Essentially, this is the idea of \emph{demonic completion} of \cite{BijlRT03}. In order to reduce the number of transitions to the chaos state, the concretization operator is parametric in a reflexive relation $\mathord{\SI}\subseteq
A\times A$ which describes whether two symbols are sufficiently similar. With this relation, we allow transitions to the chaos state only for those symbols that are not similar to any symbol for which the code is defined:

\begin{definition}[Concretization, \coqref{Concretization.v}{concretization}{In Coq, we have two equivalent definitions of the transition relation: on the one hand via the rules \eqref{rule CW1}--\eqref{rule CW4} and on the other hand as one formula.}]\label{def: concretisation}
	Let $\M\in \LTS(B)$ be an LTS, $\R \in \Code(A,B)$ an action code,
	and $\SI \subseteq A\times A$ a reflexive relation. The
	\emph{concretization} $\gamma_{\R,\SI}(\M) \in \LTS(A)$ consists of:
	\begin{itemize}
		\item $Q^{\gamma(\M)} := Q^\M \times N \cup \set{\chi}$ with
		\(
			N := \set{w\in A^*\mid w = \epsilon \text{ or }\exists b\in \dom(\R)\colon w\lneqq \R(b)}
		\).
		\item $q_0^{\gamma(\M)} := (q_0^\M,\epsilon)$
		\item Transitions are defined by the following rules:
		\[
			\infer[\rulelabel{rule CW1}{$1_\gamma$}]
				{(q,w) \xtransto[\gamma(\M)]{a} (q, wa)}{wa\in N}
			\qquad\qquad
			\infer[\rulelabel{rule CW2}{$2_\gamma$}]
				{(q,w)\xtransto[\gamma(\M)]{a} (q', \epsilon)}{q\xtransto[\M]{b}q',~~\R(b) = wa}
		\]
		\[
			\infer[\rulelabel{rule CW3}{$3_\gamma$}]
				{(q,w) \xtransto[\gamma(\M)]{a} \chi}{\forall a'\in A, (a,a')\in \SI\colon ~~ wa'\notin N \wedge wa'\notin \Im(\R)}
			\qquad\qquad
			\infer[\rulelabel{rule CW4}{$4_\gamma$}]
				{\chi \xtransto[\gamma(\M)]{a} \chi}{}
		\]
	\end{itemize}
\end{definition}

Intuitively, $N$ represent the internal nodes of the tree-representation of action code $\R$. The transitions then try to accumulate a word $w\in A^*$ known to the action code (rule \eqref{rule CW1}). As soon as we reach $w = \R(b)$ for some $b$, we use a $b$-transition in the original $\M\in \LTS(B)$ to jump to a new state (rule \eqref{rule CW2}).
The chaos state $\chi$ attracts all runs with symbols unknown to the action
code. The corresponding rule \eqref{rule CW3} involves the relation
$\SI\subseteq A\times A$. The rule only allows a transition to $\chi$ for a
symbol $a\in A$ if there is no related symbol $a'\in A$, $(a,a')\in A$ for
which the code $R$ could make a transition. For general LTSs, we can simply
consider $\SI$ to be the identity relation on $A$.
Once transitioned to the chaos state $\chi$, we allow transitions for arbitrary
action symbols $a\in A$ (rule \eqref{rule CW4}).

\begin{example}
	\label{example concretization}
	For the special case of Mealy machines $A:=I\times O$, we can define
	$\SI\subseteq (I\times O)\times (I\times O)$ to relate $(i,o)$ and $(i',o')$ iff $i=i'$, i.e.~two actions are related if they use the same input symbol. Then, we only have transitions to the chaos states if the code can't do any action for the same input symbol $i\in I$.
	\autoref{fig: example concretization} depicts the concretization (for this $\SI$) of the Mealy machine of \autoref{example contraction}(right) with the action code of Figure~\ref{example contraction}(left). To increase readability, we introduced two copies of chaos state $\chi$. Also, multiple labels next to an arrow denote multiple transitions.
	\begin{figure}[t]\centering%
		\begin{tikzpicture}[lts,node distance=2.0cm]
		\node[state, initial, initial where=above] (q0) {$q_0, \epsilon$};
		\node[state, below of=q0] (q1) {$q_2, b$};
		\node[state, right of=q1] (q2) {$q_2, \epsilon$};
		\node[state, right of=q0] (q3) {$q_0, a$};
		\node[state, left of=q0] (q0r2) {$q_0, b$};
		\node[state, left of=q1] (chi1) {$\chi$};
		\node[state, right of=q2] (q2r1) {$q_2, a$};
		\node[state, right of=q3] (chi2) {$\chi$};
		\draw[transition,overlay,bend angle=20, every node/.append style={align=center}]
		(q0) edge[bend right,swap] node{$b/0$} (q0r2)
		(q0) edge node{$a/0$} (q3)
		(q1) edge node {$a/0, a/1$} (chi1)
		(q1) edge node{$b/0$} (q0)
		(q2) edge node{$b/0$} (q1)
		(q2) edge[bend right,swap] node{$a/0$} (q2r1)
		(q3) edge node{$a/0$} (q2)
		(q3) edge node{$b/0, b/1$} (chi2)
		(q0r2) edge node[swap] {$a/0, a/1$} (chi1)
		(q0r2) edge[bend right,swap] node{$b/0$} (q0)
		(q2r1) edge node[swap] {$b/0,b/1$} (chi2)
		(q2r1) edge[bend right,swap] node{$a/0$} (q2)
		(chi1) edge[loop left] node{$a/0,a/1$,\\$b/0,b/1$} (chi1)
		(chi2) edge[loop right] node{$a/0,a/1$,\\$b/0,b/1$} (chi2)
		;
		\end{tikzpicture}
		\caption{Concretization of the Mealy machine of Figure~\ref{example contraction}.}
		\label{fig: example concretization}
	\end{figure}	
\end{example}

Like in the refinement operator,
the transition structure of $\gamma$ is built in such a way that transitions
for $b\in B$ in $\M$ correspond to runs of $\R(b)$ in $\gamma(\M)$:
\[
	(q,\epsilon) \xTransto[\gamma(\M)]{\R(b)} \bar q
	\quad\text{iff}\quad
	\exists q'\colon q\xtransto[\M]{b} q' \text{ and }\bar q = (q',\epsilon).
\]
To make $\gamma$ right adjoint to $\alpha$, all runs outside the code $\R$ lead
to the chaos state.
One may think that the many transitions to the chaos state $\chi$ would make
the construction $\gamma_\R$ trivial. However, only those paths lead to $\chi$
for which the action code is not defined.

The following technical condition describes that a code $\R$ contains sufficiently many related symbols compared to a given $\M\in\LTS(A)$:

\begin{definition}[\coqref{Concretization.v}{icomplete}{}]
	A code $\R\in \Code(A, B)$ is called \emph{$\SI$-complete for $\M \in \LTS(A)$}, if for
	all $w\in B^*$, $u\in A^*$, $q\in Q^\M$, $a,a'\in A$:
	\[
		r_0\xTransto[\R]{u\,a}\quad\text{and}\quad
		(a,a')\in \SI\quad\text{and}\quad
		q_0\xTransto[\M]{\R^*w\,u} q\xtransto{a'}
		\quad\text{ implies }\quad
		r_0\xTransto[\R]{u\,a'}.
	\]
\end{definition}
Intuitively, $\SI$-completeness means that if a state $q\in \M$ can do a
transition for $a'\in A$ which is related to similar symbol $a\in A$ defined in
the action code, then $a'\in A$ itself is also defined in the action code.
However, we do not compare arbitrary transitions of $q$ in $\M$ with arbitrary
symbols mentioned in $\R$, but only look at the node in $\R$ reached when \textqt{executing $\R$} zero or more times while following the path $q_0\xTransto{} q$.

For example, if $\SI\subseteq A\times A$ happens to be the identity relation,
then $\R$ is $\SI$-complete for any $\M\in \LTS(A)$.

\begin{assumption}
	For the rest of the present \autoref{sec: concretizations}, we fix the sets
	$A,B$, an action code $\R\in \Code(A,B)$, and a reflexive relation
	$\SI\subseteq A\times A$.
\end{assumption}

\begin{theorem}[Galois connection, \coqref{Concretization.v}{concretization_galois_connection}{In the proof of direction ($\Rightarrow$), we use the law of excluded middle for the property that for a $w\in A^*$ there is some or there is no $b\in B$ with $w\le \R(b)$; in the constructive setting, this becomes an explicit assumption.
}]
	 \label{gamma galois}
	For all $\N\in \LTS(A)$, and $\M\in \LTS(B)$, such that $\R$ is $\SI$-complete for
	$\N$, we have
	\[
		\alpha_\R(\N) \sqsubseteq \M
		~(\text{in }\LTS(B))
		\qquad\Longleftrightarrow\qquad
		\N \sqsubseteq \gamma_{\R,\SI}(\M)
		~(\text{in }\LTS(A)).
	\]
\end{theorem}

\begin{example}[\coqref{Concretization.v}{concretization_galois_connection_lts}{}]
If we instantiate $\SI$ to be the identity relation $\Delta$ on $A$, then this means that we simply replace $a'$ with $a$ in rule \eqref{rule CW3}, and then we have above equivalence for all $\N\in \LTS(A)$ and $\M\in \LTS(B)$ (without any side-condition):
\[
	\alpha_\R(\N) \sqsubseteq \M
	~(\text{in }\LTS(B))
	\qquad\Longleftrightarrow\qquad
	\N \sqsubseteq \gamma_{\R,\Delta}(\M)
	~(\text{in }\LTS(A)).
\]
\takeout{}
\end{example}
\begin{example}[\coqref{MealyExample.v}{simulation_from_mealy_to_own_concretization}{%
We have an auxiliary lemma \UScore{mealy_example_code1_is_complete} showing that $\R$ is indeed $\SI$-complete for $\N$; there is a non-trivial case, in which we verify inductively that there is no $w\in A^*$ such that $q_0 \xTransto{\R^*w} q_3$.
}]
	Consider the instantiation of $\SI$ for Mealy machines described in Example~\ref{example concretization}.
	Let $\N$ be our running example of Figure~\ref{fig:example_mealy_machine},
	let $\R$ be the action code from Figure~\ref{example contraction}(left), and
	let $\M$ be the abstract Mealy machine from Figure~\ref{example contraction}(right), i.e.\
	$\alpha_\R(\N) = \M$.
	One can verify that $\R$ is $\SI$-complete for $\N$.\fvnote{}
	Therefore, application of the Galois connection gives that there is a simulation from $\N$ to the Mealy machine $\gamma_{\R, \SI}(\M)$ of Figure~\ref{fig: example concretization}.
\end{example}

It is a standard proof that the operators in a Galois connections are monotone.
In that proof, one applies the Galois connection also to
$\M:=\gamma_{\R,\SI}(\N)$, so we first need to show that it satisfies the
technical completeness condition:
\begin{lemma}[\coqref{Concretization.v}{icomplete_for_concretization}{}]
	$\R$ is always $\SI$-complete for $\gamma_{\R,\SI}(\M)$.
\end{lemma}

\begin{corollary}[\coqref{Concretization.v}{concretization_monotone}{}] \label{gamma monotone}
	$\M\sqsubseteq \N$ in $\LTS(B)$ implies
	$\gamma_{\R,\SI}(\M)\sqsubseteq \gamma_{\R,\SI}(\N)$ in $\LTS(A)$.
\end{corollary}

\begin{iflong}
\begin{remark}
	Monotonicity of concretization also follows by observing that the rules in Definition~\ref{def: concretisation} all fit the \emph{tyft} format of \cite{GrV92} if we view $( \cdot, w)$ as a unary operator for each sequence $w \in N$. Monotonicity then follows from the result of \cite{GrV92} that the simulation preorder is a congruence for any operator defined using the \emph{tyft} format.
	Since contraction also can be defined using the \emph{tyft} format, also monotonicity of contraction (Proposition~\ref{contraction monotone}) follows from the result of \cite{GrV92}.
\end{remark}
\end{iflong}

Like refinement, concretization preserves determinism.

\begin{proposition}[\coqref{Concretization.v}{concretization_preserves_determinism}{}] \label{gamma deterministic}
	If $\M\in \LTS(B)$ is a deterministic LTS and $\Delta$ the identity relation on $A$, then $\gamma_{\R,\Delta}(\M)$ is
	deterministic, too.
\end{proposition}

If the code $\R\in \Code(A,B)$ is defined for all labels mentioned in $\M\in
\LTS(B)$, then $\gamma_\R$ is even the right inverse of $\alpha_\R$, that is,
we have a Galois insertion:
\begin{theorem}[Galois insertion, \coqref{Concretizaton.v}{concretization_galois_insertion}{}]\label{gamma galois insertion}
	If $\M\in \LTS(\dom(\R))$, then $\M\cong \alpha_\R(\gamma_{\R,\SI}(\M))$.
\end{theorem}
\noindent
Note that $\dom(\R)\subseteq B$, and so $\LTS(\dom(\R))\subseteq \LTS(B)$.
Since we may reach the chaos state $\chi$ in the concretization, it is clear that $\gamma_\R$ is not a left inverse of $\alpha_\R$ in general.

\section{Action Code Composition}
\label{sec: composition}
Since notions of abstraction can be stacked up, it is natural to consider
multiple adaptors for multiple action codes.
Assume an action code $\R\in \Code(A,B)$ and an action code $\S\in \Code(B,C)$. Then the composition of $\R$ and $\S$ should be an action code from $A$ to $C$.

\begin{definition}[\coqref{CodeMap.v}{compose_codemap}{}]
	Given two map-based action codes $\R\colon B\partialto A^+$ and $\S\colon
	C\partialto B^+$, we define their \emph{(Kleisli) composition}
	$(\R * \S) \colon C\partialto A^+$
	by
	\twnote{}
	\[
		(\R * \S)(c) = \begin{cases}
			\R(b_1)\cdots \R(b_n)
			&\text{if }\S(c) = b_1\cdots b_n
			\text{ with }\forall i\colon b_i \in \dom(\R)
			\\
			\text{undefined}
			&\text{otherwise}
		\end{cases}
	\]
\end{definition}
The composed action code $\R*\S$ is only defined for $c\in C$ if $\S$ is
defined for $c$ and additionally $\R$ is defined for every letter $b_i\in B$
that appears in the word $\S(c) \in B^+$.
\begin{remark}\label{remark kleisli}
	The defined composition is an instance of \emph{Kleisli composition} for a monad,
	which is a standard concept in functional programming and
	category theory.
\end{remark}

\begin{lemma}[\coqref{CodeMap.v}{compose_code}{}]\label{lem:composition}
	Action codes are closed under composition.

	Concretely, given two map-based action codes $\R\colon B\partialto A^+$ and
	$\S\colon C\partialto B^+$, their Kleisli composition
	$(\R * \S) \colon C\partialto A^+$
	is again a \emph{prefix-free} partial map.
\end{lemma}

Now that we can compose action codes, we can now investigate how the previously defined operators on LTSs behave for composed action codes:

\begin{theorem}[\coqref{Contraction}{contraction_code_composition}{Item for $\alpha$},
\coqref{RefinementCodeComposition}{refinement_preserves_code_composition}{Item for $\rho$.}]\label{thm code composition}
	Contraction and refinement commute with action code composition: for action codes $\R\in
	\Code(A,B)$, $\S\in \Code(B,C)$,
	\begin{enumerate}
	\item\label{alpha code composition}
		\(
			\alpha_{\R*\S}(\M) = \alpha_{\S}(\alpha_{\R}(\M))
		\) for all $\M\in \LTS(A)$.
	\item\label{refinement code composition}
		\(
			\rho_{\R*\S}(\M) = \rho_{\R}(\rho_{\S}(\M))
		\), whenever $\Im(\S) \subseteq \dom(\R)^+$ and
		for all $\M\in \LTS(C)$.
	\end{enumerate}
\end{theorem}
For the case of refinement, the additional assumption expresses that every word produced by $\S$ only
contains letters $b\in B$ for which $\R$ is defined. The equations of \autoref{thm code composition} equivalently mean that the following diagrams commute:
\[
		\begin{tikzcd}[row sep=3mm]
			\LTS(A)
			\arrow{dr}[swap]{\alpha_\R}
			\arrow{rr}{\alpha_{\R*\S}}
			& & \LTS(C)
			\\
			& \LTS(B)
			\arrow{ur}[swap]{\alpha_\S}
		\end{tikzcd}
		\qquad
		\begin{tikzcd}[row sep=3mm]
			\LTS(A)
			\arrow[<-]{dr}[swap]{\rho_\R}
			\arrow[<-]{rr}{\rho_{\R*\S}}
			& & \LTS(C)
			\\
			& \LTS(B)
			\arrow[<-]{ur}[swap]{\rho_\S}
		\end{tikzcd}
\]

\begin{remark}[\coqref{Contraction.v}{contraction_does_not_preserve_code_composition}{In \autoref{appendix} on page \pageref{details contraction composition}, we visualize the involved LTSs for the counter example. In the Coq formalization, we directly show that the existence of an isomorphism leads to a contradiction.}]
\label{concretization no code composition}
	Concretization does not commute with action code composition. The reason for that is that the rules \eqref{rule CW1} and \eqref{rule CW2} in $\gamma_{\R}(\gamma_{\S}(\M))$ would also be applied to transitions for the chaos state in $\gamma_{\S}(\M) \in \LTS(B)$
	(see appendix for details).\twnote{}
\end{remark}

\section{Adaptors}
\label{sec: adaptors}
In this section, we describe how action codes may be used for learning and testing of black-box systems.  The general architecture is shown in Figure~\ref{fig:architecture}. On the right we see the \emph{system under test (SUT)}, some piece of hardware/software whose behavior can be modeled by a Mealy machine $\M$ with inputs $I$ and outputs $O$.  
\begin{figure}[thb]%
	\begin{minipage}{.7\textwidth}%
	\centering%
		\begin{tikzpicture}[
			x=35mm, %
			y=35mm,
			big node/.style={
				align=center,
				shape=rectangle,
				draw=lipicsYellow,
				line width=1pt,
				rounded corners=1pt,
				font=\small,
				minimum width=20mm,
				minimum height=25mm,
			}]
			\node[big node] (learner) at (0,0) {Learner /\\ Tester};
			\node[big node] (adaptor) at (1,0) {Adaptor \\ $\R$};
			\node[big node] (sut) at (2,0) {SUT \\ $\M$};
			\def\adaptorEdgeOffset{1cm}
			\path[-stealth,auto,
			shorten >= 1mm,
			shorten <= 1mm,
			shift up/.style={
				transform canvas={yshift=4mm},
			},
			shift down/.style={
				transform canvas={yshift=-4mm},
			},
			]
				(learner) edge[shift up] node {$x\in X$} (adaptor)
				(adaptor) edge[shift up] node {$i\in I$} (sut)
				(sut) edge[shift down] node {$o\in O$} (adaptor)
				(adaptor) edge[shift down] node {$y\in Y$} (learner)
			;
		\end{tikzpicture}%
	\caption{Using action codes for learning/testing.}
	\label{fig:architecture}
	\end{minipage}\hfill%
	\begin{minipage}{.28\textwidth}\quad%
		\begin{tikzpicture}[lts,empty states,x=8mm,y=12mm,
		every label/.append style={
			text depth=0, %
		},
		]
		\node[initial above, state] (0) at (0,1) {\hidestatename{$r_0$}};
		\node[state, label={below:$\mathbf{0}/A$}] (1) at (-1,0) {\hidestatename{$r_1$}};
		\node[state, label={below:$\mathbf{0}/B$}] (2) at (1,0) {\hidestatename{$r_2$}};
		
		\path[transition]
		(0) edge node[swap] {a/0} (1)
		edge node {b/0} (2)
		;
		\end{tikzpicture}
		\caption{\autoref{exNotDeterminiate}}%
		\label{example indeterminate}
	\end{minipage}%
\end{figure}
On the left we see the \emph{learner/tester}, an agent which either tries to construct a model $\N$ of $\M$ by performing experiments, or already has such a model $\N$ and performs experiments (tests) to find a counterexample which shows that $\M$ and $\N$ behave differently.  The learner/tester uses abstract inputs $X$ and outputs $Y$.
In between the learner/tester and the SUT we place an \emph{adaptor}, which uses action code $\R$ to translate between the abstract world of the learner/tester and the concrete world of the SUT.
In order to enable the adaptor to do its job, we need to make four (reasonable) assumptions.

Our first assumption, common in model-based testing \cite{Tret96}, is that the SUT will accept any input from $I$ in any state, that is, we require that $\M$ is \emph{input enabled}: for all $q \in Q^\M$ and $i \in I$, $q \xtransto{i/}_\M$.
Our second assumption is that code $\R$ is $\SI$-complete for $\M$. This ensures that whenever the adaptor sends a concrete symbol $i \in I$ to the SUT, it will accept any output $o \in O$ that the SUT may possibly produce in response.
Our third assumption is that whenever the adaptor receives an abstract input $x \in X$ from the learner/tester, it can choose concrete inputs from $I$ that drive $\R$ from its initial state to a leaf with label $(x, y)$, for some $y \in Y$.  Output $y$ can then be returned as a response to the learner/tester. Reaching such a leaf is nontrivial since the transitions taken in $\R$ are also determined by the outputs provided by the SUT. We may think of the situation in terms of a 2-player game where the adaptor wins if the game ends in an $x$-leaf, and the SUT wins otherwise. Formally, we require that $\R$ has finitely many states and a winning strategy for every input $x \in X$, as defined below:

\begin{definition}[Winning]
	Let $\R = \tuple{R,r_0,\transto, l} \in \Code(I\times O, X\times Y)$ be an action code with $R$ finite and let $x \in X$. Then
	\begin{enumerate}
		\item 
		A leaf $r \in R$ is \emph{winning for} $x$ if $\pi_1 (l(r)) = x$.\footnote{We use projections functions $\pi_1$ and $\pi_2$ to denote the first and second element of a pair, respectively. So $\pi_1(x,y) = x$ and $\pi_2(x, y) = y$.}
		\item
		An internal state $r \in R$ is \emph{winning for $x$ with input $i \in I$} if $r \xtransto{i/}$ and, for each transition of the form $r \xtransto{i/o} r'$, $r'$ is winning for $x$. 
		\item
		An internal state $r \in R$ is \emph{winning for $x$} if it is winning for $x$ with some $i \in I$. 
		\item 
		$\R$ has a \emph{winning strategy for} $x$ if $r_0$ is winning for $x$.
	\end{enumerate}
\end{definition}

\begin{example}
	The action codes for Mealy machines that we have seen thus far (Figures~\ref{D4}, \ref{example contraction} and \ref{example contraction2}) are winning for all the inputs that label their leafs.
	The action code of Figure~\ref{D4} is not winning for the input \includegraphics[scale=0.04]{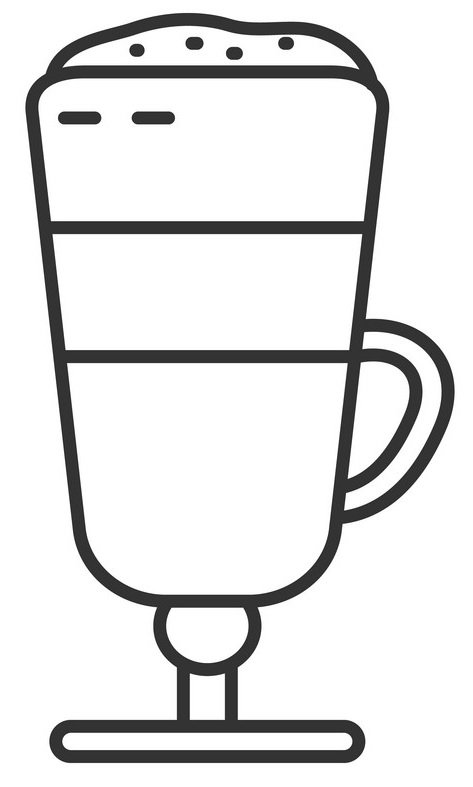} (latte macchiato), for the simple reason that this input does not label any leaf. If we remove the transition to the leaf \includegraphics[scale=0.15]{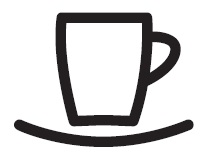}$/2$ in Figure~\ref{D4}, then the resulting code is no longer winning for \includegraphics[scale=0.15]{espresso.jpg} (espresso), although it is winning for \includegraphics[scale=0.15]{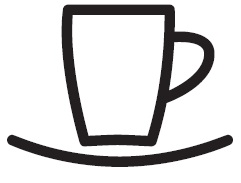} (coffee).
\end{example}

Our fourth and final assumption is that action code $\R$ is \emph{determinate}.  If an action code is determinate then, for each state $r$ and abstract input $x$, there is at most one concrete input $i$ such that $r$ is winning for $x$ with $i$.

\begin{definition}[Determinate, \coqref{Adaptor.v}{Determinate}{%
In the formalization, we define \emph{determinate} for general codes $\R\in \Code(A,B)$ and model that two action symbols $a_1,a_2\in A$ having the same input component by a relation $\SI_A\subseteq A\times A$ -- in the same style as for the concretization operator. Likewise, we consider a relation $\SI_B\subseteq B\times B$ that models that two $b_1,b_2\in B$ have the same input component.
}]
	An action code $\R$ is \emph{determinate} if, for each state $r$, whenever $r \xtransto{i_1/} r_1$, $r \xtransto{i_2/} r_2$ and from both $r_1$ and $r_2$ there is a path to a leaf labeled with input $x$, then $i_1 = i_2$.
\end{definition}

\begin{example} \label{exNotDeterminiate}
	All action codes for Mealy machines that we have seen thus far (Figures~\ref{D4}, \ref{example contraction} and \ref{example contraction2}) are determinate.
	\autoref{example indeterminate} shows an action code that is not determinate: in the root two different concrete inputs $a$ and $b$ are enabled that lead to leafs with the same abstract input $\mathbf{0}$. Hence (trivially), this action code does have a winning strategy for input $\mathbf{0}$.
\end{example}

Algorithm~\ref{alg: adaptor} shows pseudocode for an adaptor that implements action code $\R$.
During learning/testing, the adaptor records the current state of the action code in a variable $r$.
When an abstract input $x$ arrives, it first sets $r$ to $r_0$.
As long as current state $r$ is internal, the adaptor chooses an input $i$ that is winning for $x$, and forwards it to the SUT.  
When the SUT replies with an output $o$, the adaptor sets $r$ to a state $r'$ with $r \xrightarrow{i/o} r'$.
When the new $r$ is internal the adaptor chooses again a winning input, and updates its current state after interacting with the SUT, etc.
When the new $r$ is a leaf with label $(x, y)$ then the adaptor returns symbol $y$ to the learner/tester and waits for the next abstract input to arrive. 
\begin{algorithm}[ht!]
	\begin{algorithmic}[1]
		\While{$\mathit{true}$}
		\State $x \gets \code{Receive-from-learner}()$ 
		\State $r \gets r_0$ 
		\While{$r$ is internal} \Comment{loop invariant: $r$ is winning for $x$}
		\State $i \gets$ unique input such that $r$ is winning for $x$ with $i$
		\State $\code{Send-to-SUT}(i)$ 
		\State $o \gets \code{Receive-from-SUT}()$
		\State $r \gets$ unique state $r'$ such that $r \xtransto{i/o} r'$ \Comment{$\R$ is $\SI$-complete for $\M$}
		\EndWhile	
		\State $\code{Send-to-learner}(\pi_2(l(r)))$	
		\EndWhile
	\end{algorithmic}%
	\caption{Pseudocode for an adaptor that implements %
		action code $\R$. 
	}
	\label{alg: adaptor}
\end{algorithm}

From the perspective of the learner/tester, the combination of the adaptor and SUT behaves the same as the contraction $\alpha_\R(\M)$. 
In the appendix, we will formalize this statement by modeling both the combination of adaptor and SUT, as well as contraction $\alpha_\R(\M)$ as expressions in the process calculus CCS \cite{Mi89}, and then establish the existence of \emph{delay simulations} between these expressions.
This implies that both expressions have the same traces if we remove all occurrences of the synchronizations between adaptor and SUT, which are invisible from the perspective of the learner.

\begin{theorem}\label{alpha implementation}
	Let $\M\in \LTS(I\times O)$ be an input enabled Mealy machine and let
	$\R \in \Code(I \times O, X \times Y)$ be a finite, determinate action code that has a winning strategy for every input in $X$ and that is output enabled for $\M$.
	Then the composition of an implementation for $\M$ and an adaptor for $\R$ is delay simulation equivalent to an implementation for $\alpha_\R(\M)$.
\end{theorem}

Active automata learning algorithms and tools for Mealy machines typically assume that the system under learning is \emph{output deterministic}\footnote{The notion of deterministic that we use in this article is the standard one for LTSs. In the literature on Mealy machines and FSMs, machines that we call output deterministic are called deterministic, and machines that we call deterministic are called observable.}: the output and target state of a transition are uniquely determined by its source state and input.

\begin{definition}
	Mealy machine $\M$ is \emph{output deterministic} if, for each state $q$ and input $i$,
	\begin{eqnarray*}
		q \xtransto{i / o} r \wedge q \xtransto{i / o'} r' & \Rightarrow & o=o' \wedge r=r'.
	\end{eqnarray*}	
\end{definition}
For action codes that are determinate, contraction preserves output determinism.  This property makes it possible to use existing automata learning tools to learn models of an output deterministic SUT composed with a determinate adaptor.

\begin{proposition}[\coqref{Adaptor.v}{contraction_preserves_determinism}{%
	The formalized property \emph{determinate} is generic in two relations $\SI_A \subseteq A\times A$ and $\SI_B\subseteq B\times B$, and likewise, the formalization of this proposition uses a notion of determinism generic in such a relation $\SI$.
	Thus, we obtain the proposition in the paper as a special case for the relations $\SI_A$ resp.\ $\SI_B$ modeling same input (cf.~\autoref{example concretization}).~
	As another instance, we obtain that contraction $\alpha_\R$ for any code $\R$ sends deterministic LTSs to deterministic LTSs.
}]
	\label{La preservation properties by contraction}
	Suppose $\M$ is a Mealy machine and $\R$ is an action code.
	If $\M$ is output deterministic and $\R$ is determinate then $\alpha_\R(\M)$ is output deterministic.
\end{proposition}

\section{Discussion and Future Work}
\label{sec: discussion}

Via the notion of action codes, we provided a new perspective on the fundamental question how high-level state machine models with abstract actions can be related to low-level models in which these actions are refined by sequences of concrete actions.
This perspective may, for instance, help with the systematic design of adaptors during learning and testing, and the subsequent interpretation of obtained results.
Our theory allows for action codes (such as in Figure~\ref{D4}) that are adaptive in the sense that outputs which occur in response to inputs at the concrete level may determine the sequence of concrete inputs that refines an abstract input. We are not aware of case studies in which such adaptive codes are used, but believe they may be of practical interest.  One may, for instance, consider a scenario in which an abstract action AUTHENTICATE is refined by a protocol in which a user is either asked to authenticate by entering a PIN code, or by providing a fingerprint.

Close to our work are the results of Rensink and Gorrieri \cite{RensinkG01}, who investigate vertical imple\-men\-tation relations to link models at conceptually different levels of abstraction. These relations are indexed by a refinement function that maps abstract actions into concrete processes.  Within a setting of a CCS-like language, Rensink \& Gorrieri \cite{RensinkG01} list a number of proof rules that should hold for any vertical implementation relation, and propose \emph{vertical bisimulation} as a candidate vertical implementation relation for which these proof rules hold.
In the setting of our paper, we can define two vertical implementation relations $\sqsubseteq^\R_\gamma$ and $\sqsubseteq^\R_\rho$, for any action code $\R$, by
\begin{eqnarray*}
	\M  \sqsubseteq^\R_\gamma \N ~ \Leftrightarrow ~ \M \sqsubseteq \gamma_\R (\N) & \mbox{ and } &
	\M  \sqsubseteq^\R_\rho \N ~  \Leftrightarrow ~ \M \sqsubseteq \rho_\R (\N).
\end{eqnarray*}
Then $\mathord{\sqsubseteq^\R_\rho} \subseteq \mathord{\sqsubseteq^\R_\gamma}$ and both relations satisfy all  language-independent proof rules of \cite{RensinkG01}.
\begin{iflong}
For instance, we have
\[
\infer{\M  \sqsubseteq^\R_\gamma \N}{\M \sqsubseteq \M'& \M'  \sqsubseteq^\R_\gamma \N' & \N'\sqsubseteq \N} 
\]
(since $\gamma_\R$ is monotone and $\sqsubseteq$ is transitive). 
\end{iflong}
With the action code $\R$ of Figure~\ref{ascii}, both implementation relations relate the LTSs of Figures~\ref{fig:desired refinement} and \ref{ab}.
However, the vertical bisimulation preorder of Rensink and Gorrieri \cite{RensinkG01} does not relate these LTSs, when using a code that maps $a$ to $1 \, 4 \, 1$, and $b$ to $1 \, 4 \, 2$.
This suggests that bisimulations may not be suitable as vertical implementation relations.

Also close to our work are results of Burton et al \cite{BurtonKP01,KP98}, who propose a vertical implementation relation in the context of CSP.  Instead of action codes, they use \emph{extraction patterns}, a strict monotonic map $\mathit{extr} : \mathit{Dom} \rightarrow B^*$, where $\mathit{Dom}$ is the prefix closure of a set $\mathit{dom} \subseteq A^*$ of concrete action sequences that may be regarded as complete. As a mapping from concrete to abstract sequences of actions, extraction patterns are more general than action codes. %
However, as extraction mappings are not required to have an inverse, establishing interesting  Galois connections in this setting may be difficult.
With an extraction pattern defined in the obvious way, the LTSs of Figures~\ref{fig:desired refinement} and \ref{ab} are related by the implementation relation of \cite{BurtonKP01}.
We are not aware of any other vertical implementation relation proposed in the literature that handles our basic interface refinement example correctly. 
\begin{iflong}
	We find it surprising that the fundamental problem of refining inputs actions has not been properly addressed in the literature, except in some work that apparently has not been picked up outside Newcastle-upon-Tyne and Catania.
\end{iflong}

Our theory is orthogonal to the one of Aarts et al \cite{AJUV15}, which explores the use of so-called \emph{mappers} to formalize adaptors that abstract the large action alphabets of realistic applications into small sets of actions that can be handled by a learning tool.  Aarts et al \cite{AJUV15} also describe the relation between abstract and concrete models using a Galois connection.  In practical applications of model learning, it makes sense to construct an adaptor that combines a mapper in the sense of \cite{AJUV15} with an action code as introduced in this paper. Fiter\u{a}u-Bro\c{s}tean et al \cite{FJV16} describe a small domain specific language to specify mapper components, and from which adaptor software can be generated automatically. It would be interesting to extend this domain specific language so that it may also be used to specify action codes.

We developed our theory for LTSs and Mealy machines, using the simulation preorder as the implementation relation.  It would be interesting to transfer our results to other modeling frameworks, such as IOTSs \cite{Tret96} timed automata \cite{Uppaal4.0} and Markov Decision Processes, and to other preorders and equivalences in the linear-time branching-time spectrum for LTSs \cite{vG01} and IOTSs \cite{JanssenVT19}.
An obvious direction for future work would be to explore how action codes interact with parallel composition. Here the work of \cite{BurtonKP01,KP98} may serve as a basis.

Different action codes lead to different contractions, and thereby to different abstract views of a system, see for instance Figures~\ref{example contraction} and~\ref{example contraction2}.  We may try to exploit this fact during learning and testing. For instance, if a system $\M$ is too big for state-of-the-art learning algorithms, we may still succeed to learn partial views using cleverly selected action codes. Using our Galois connections we then could obtain various upper and lower bounds for $\M$. Ideally, such an approach may even succeed to uniquely identify $\M$.
\begin{iflong}
Maarse \cite{Maa20} quantified the quality of a contraction $\alpha_\R(\M)$ in terms of the graph-theoretic concept of \emph{eccentricity}.  If $q$ and $q'$ are states in an LTS $\M$ then $d(q, q')$ is defined as the number of transitions in the shortest path from $q$ to $q'$ (or $\infty$ if no such path exists).
For any set of states $Q \subseteq Q_\M$, the \emph{eccentricity} $\epsilon(Q)$ is defined as
$\max_{q' \in Q_\M} ~ \min_{q \in Q} ~ d(q, q')$,
that is, the maximal distance one needs to travel to visit a state of $\M$, starting from a state of $Q$.  A good contraction has a small set of states $Q$ and a low eccentricity $\epsilon(Q)$: it only covers a small subset $Q$ of the states of $\M$, but any state from $\M$ can be reached via a few transitions from a $Q$-state.
\end{iflong}

\label{maintextend} %
\bibliographystyle{plain}
\bibliography{references,abbreviations,dbase}

\clearpage
\appendix
\section{Formalization in Coq}
\label{coqtoc}%
\ifthenelse{\boolean{anonymous}}{%
The full Coq formalization can be accessed anonymously via:
\begin{quote}
\url{https://}
\end{quote}
}{%
The full Coq formalization is available in the ancillary files on arxiv:
\begin{quote}
\url{https://arxiv.org/src/2301.00199/anc}
\end{quote}
}
This repository contains both the plain Coq source files and generated HTML
documentation for easier reading and navigation (clickable links, etc). We use
the Coq version 8.16.1. In order to compile the source files, run
\begin{quote}
\texttt{coq\_makefile -o Makefile *.v} \\
\texttt{make}
\end{quote}
in the source directory. The HTML documentation can be read \emph{without} compilation.

For each definition or result that was formalized in Coq, the following list
provides the identifier of the corresponding object in Coq. In some places the
formalization slightly differs from what we use in the paper; if this is the
case, we also provide explanations here:

\printcoqreferences

\section{Omitted Proofs and Additional Details}%
\label{appendix}%
\nopagebreak%
\begin{proofappendix}{treeCodeToMap}
	Take equation \eqref{defMapR} as the definition of $f$:
	\[
		f(b) = \begin{cases}
			w & \text{if there are }r\in L, w\in A^*\text{ with }
			    r_0\xTransto[\R]{w} r, l(r) = b,
					\\
			\text{undefined} & \text{otherwise}.
		\end{cases}
	\]
	Since $r_0\notin L$, the range of $f$ indeed restricts to $A^+\subsetneqq A^*$.
	For well-definedness of $f$, first note that due to the injectivity of the labeling $l\colon L\to B$,
	there is at most one $r\in L$ with $l(r) = b$, and by $\R$ being tree-shaped,
	there is precisely one path $r_0\xTransto[\R]{} r$.
	Let us now prove the required properties of this partial map:
	\begin{itemize}
	\item \textbf{Prefix-freeness \eqref{prefix-free}.}
	Consider $b,b'\in B$
	with $f(b) \le f(b')$. Hence, we have runs
	\[
		r_0 \xTransto[\R]{w} r\quad r\in L\quad l(r) = b\quad w = f(b)
	\]
	and
	\[
		r_0 \xTransto[\R]{w'} r'\quad r'\in L\quad l(r') = b' \quad w' = f(b').
	\]
	By $f(b) \le f(b')$, there is some $u\in A^*$ such that $wu =w'$. Thus, the
	run for $w'=f(b')$ can be decomposed for some $\bar r\in Q^\R$:
	\[
		r_0 \xTransto[\R]{w} \bar r \xTransto[\R]{u} r'.
	\]
	The LTS $\R$ is required to be deterministic, so $\bar r = r$.
	Since $r\in L$ is a leaf, i.e.~a dead-lock state, we necessarily have $u=\epsilon$ and $r=r'$.
	Finally, $b = l(r) = l(r') = b'$.

	\item \textbf{Characterizing property \eqref{defMapR}.}
	Verification is immediate because both the property \eqref{defMapR} and the
	definition $r$ use the same witnesses, and every witness $w\in A^*$ must be a
	non-empty word.

	\item \textbf{Uniqueness.}
	Consider another prefix-free partial map $g\colon B\partialto A^+$ satisfying
	\begin{equation*}
		\text{for all }b\in B, w\in A^+\colon\qquad g(b) = w
		\quad\text{ iff }\quad
		\exists r \in L\colon
		r_0\xTransto[\R]{w} r,~~ l(r) = b.
	\end{equation*}
	Then, it is immediate that $\dom(g) = \Im(L) = \dom(f)$ and that $g(b) = f(b)$ for all $b\in \Im(L)$.
	\qedhere
	\end{itemize}
\end{proofappendix}

\begin{proofappendix}{mapCodeToTree}
	Define action code $\R$ as follows:
	\begin{itemize}
	\item $R := \set{w\in A^*\mid w=\epsilon\text{ or }\exists b\in \dom(f)\colon w\le f(b)}$,
	\item $r_0 := \epsilon$,
	\item we put $v \xtransto[\R]{a} w$ iff there is $a\in A$ with $va = w$,
	\item $l(w)$ is the unique $b\in B$ with $f(b) = w$.
	\end{itemize}
	Let us first verify that this is a well-defined action code:
	\begin{itemize}
	\item There is nothing to be verified about the state set $R$ (note that we do not require finiteness in the definition of action code).
	\item By definition of $R$, the initial state $r_0:=\epsilon$ is an element of $R$.
	\item For the transition structure we need to verify several properties:
		\begin{itemize}
		\item It is grounded because for every $w\in A^*$ with $w\le f(b)$, $b\in \dom(f)$, the state $f(b)$ is the witnessing leaf (i.e.~dead-lock state) below $w$.
		\item It is deterministic: whenever we have $v\xtransto[\R]{a} w$ and $v\xtransto[\R]{a} w'$, we obtain $w = va = w'$.
		\item It is tree-shaped: for each $w\in R$, there is a unique run to it, namely $r_0\xTransto[\R]{w} w$.
		\end{itemize}
	\item Define set $L \subseteq R$ by $L = \set{f(b) \mid b\in B}$.
	Then $L$ does not contain the initial state $r_0$, but every leaf $w\in Q \setminus \{ r_0 \}$ is in $L$. Conversely, every $w\in L$ is a leaf because $f$ is prefix-free \eqref{prefix-free}.

	Likewise, $l\colon L\to B$ is injective because $f$ is prefix-free
	\eqref{prefix-free}.
	\end{itemize}

	For uniqueness, consider another tree-shaped action code $\bar \R$ with set of non-root leaves $\bar L$ satisfying \eqref{defMapR}, concretely
	\begin{equation}
		\text{for all }b\in B, w\in A^+\colon
		f(b) = w
		\quad\text{iff}\quad
		\exists \bar r \in \bar L \colon r_0\xTransto[\bar \R]{w} \bar r
		\quad
		l^{\bar \R}(\bar r) = b.
		\label{otherCharact}
	\end{equation}
	We now need to establish an isomorphism $\phi\colon \bar \R\to \R$. Define the map $\phi\colon R^{\bar \R}\to R^{\R}$ by
	\[
		\phi(\bar r) = \text{the unique $w\in A^*$ with } r_0^{\bar R}\xTransto[\bar \R]{w} \bar r .
	\]
	This map is well defined because $\bar \R$ is tree-shaped. To see that
	$\phi(\bar r) \in A^*$ is also in $R^{\R}\subseteq A^*$ for every state $\bar r$ of $\bar
	\R$, let $\bar r' \in \bar L$ be an arbitrary leaf with $\bar r \xTransto[\bar \R]{u} \bar r'$
	in $\bar \R$ -- such a leaf exists because every action code $\bar \R$ is
	required to be grounded. Thus, for $b=l^{\R}(\bar r' )$ we have $\phi(\bar r) \le wu =
	f(b)$ by
	\eqref{otherCharact}, and so $\phi(\bar r) \in R^\R$ by above definition of $\R$.
	With our definition of $\phi$ and $\xtransto[\R]{}$, it is mechanical to check that
	$q\xtransto{a}q'$ iff $\phi(q)\xtransto{a}\phi(q')$.
	Since $\bar \R$ is tree-shaped, $\phi$ is injective. For surjectivity, consider $w\in R^{\R}$.
	If $w = \epsilon$ then we have $\phi (r_0^{\bar R}) = \epsilon$.
	Otherwise, pick any $b\in \dom(f)$ with $w\le f(b)$.
	By \eqref{otherCharact}, we have $\bar r\in \bar L$ with $q_0^{\bar \R}\xTransto[\bar
	\R]{f(b)} r$ in $\bar \R$. The intermediate state $\bar r'$ with
	\[
		q_0^{\bar \R}\xTransto[\bar \R]{w} \bar r'
		\xTransto[\bar \R]{u} \bar r
		\qquad w \, u = f(b)
	\]
	does satisfy $\phi(\bar r') = w$. Thus, $\phi$ is surjective and bijective in total.

	Also, the labelling is preserved by $\phi\colon \bar \R\to \R$ because for
	every leaf $\bar r \in \bar L$ with
	\[
		r_0^{\bar \R} \xTransto[\bar \R]{w} \bar r
	\]
	we have $f(l^{\bar \R}(\bar r)) = w$ by \eqref{otherCharact}, and so 
	\[
		l(\phi(\bar r)) = l(w) \overset{\text{def }l}{=} l^{\bar \R}( \bar r).
	\]
	because $b:=l^{\bar \R}( \bar r)$ satisfies $f(b) = w$.
\end{proofappendix}

\begin{proofappendix}{contraction monotone}
	Given a simulation $T$ from $\M$ to $\N$, we will show that $S := T \cap
	(Q^{\alpha(\M)} \times Q^{\alpha(\N)})$ is a simulation relation from
	$\alpha_\R (\M)$ to $\alpha_\R (\N)$; in other words $S\subseteq
	Q^{\alpha(\M)}\times Q^{\alpha(\N)}$ is the restriction of $T\subseteq
	Q^{\M}\times Q^{\N}$ to the subsets $Q^{\alpha(\M)}\subseteq Q^\M$ and
	$Q^{\alpha(\N)}\subseteq Q^\N$.

	For the initial states, we directly have
	\(
		(q^{\alpha(\M)}_0, q^{\alpha(\N)}_0) \in S
	\)
	because 
	\[
		(q^{\alpha(\M)}_0, q^{\alpha(\N)}_0)
		= (q^{\M}_0, q^{\N}_0)
		\in T.
	\]
	For transitions,
	consider $(q,p) \in S$ and $q\xtransto{b} q'$ in $\alpha(\M)$. By definition of $\alpha(\M)$, we have
	\[
		b\in \dom(\R)\text{ and }
		q\xTransto{\R(b)} q'
		\qquad\text{in }\M.
	\]
	In $\M$ and $\N$, the states $(q,p)\in S\subseteq T$ are also related by the simulation, and so we obtain
	\[
		p\xTransto{\R(b)} p'
		\qquad\text{in }\N
		\text{ with }(q',p')\in T.
	\]
	By the definition of $\alpha$, we have $p'\in Q^{\alpha(\N)}$ and $p\xtransto{b} p'$ in $\alpha(\N)$.
	Thus, also $(q',p')\in S$ as desired.
\end{proofappendix}

\begin{example}[Trace semantics]
	\label{alphaTrace}
	The trace semantics $\obs$ is closely related to concretization.
	Define $\delta\colon \LTS(A)\to \LTS(A+\set{\$})$ by
	$\delta((Q,q_0,\transto)) = (Q+ \set{\$}, q_0, \mathord{\transto}\cup \set{(q,\$)\mid
	q\in Q})$ and define the action code $\R\in \Code(A+\set{\$}, B)$ to $B:=A^*$
	such that the concrete word $a_1\cdots a_n\$$ is related to the abstract
	symbol $a_1\cdots a_n \in B$, then $\alpha_\R\circ \delta\colon \LTS(A)\to
	\LTS(A^*)$ sends every $\M\in \LTS(A)$ to a system in $\LTS(A^*)$ with states $\set{q_0,\$}$
	and transitions
	\[
		q_0\xtransto{w} \$
		\quad
		\Longleftrightarrow
		\quad
		w\in \trace(\M).
	\]
\end{example}

\begin{proofappendix}{rhoMonotone}
	Let $S\subseteq Q^\M\times Q^\N$ be the simulation witnessing $\M \sqsubseteq \N$.
	Let us verify that
	\[
		T := \set{((q,w),(p,w')) \in Q^{\rho(\M)}\times Q^{\rho(\N)}\mid 
		w = w', (q,p)\in S}
	\]
	is a simulation from $\rho(\M)$ to $\rho(\N)$.
	Clearly,
	\[
		q_0^{\rho(\M)}
		= (q_0^{\M}, \epsilon)
		\mathrel{T} (q_0^{\N}, \epsilon)
		= q_0^{\rho(\N)}.
	\]
	Consider a pair in $T$, i.e.~$(q,w)$, $(p,w)$ with $q\mathrel{S}p$:
	\begin{enumerate}
	\item For transitions produced by rule \eqref{rho end},
		\[
			(q,w)\xtransto[\rho(\M)]{a} (q',\epsilon),
		\]
		we have $q\xtransto[\M]{b} q'$ for some $b\in\dom(\R)$ with $\R(b) = wa$ by
		rule assumption.
		Since $(q,p) \in S$, the simulation $S$ provides us with a transition
		\[
			p\xtransto[\N]{b} p'
			\qquad\text{in }\N
			\qquad\text{and}\qquad
			(q',p')\in S.
		\]
		Thus, we can apply rule \eqref{rho end} in $\rho(\N)$ to obtain
		\[
			(p,w)\xtransto[\rho(\N)]{a} (p',\epsilon)
			\qquad\text{in }\rho(\N)
		\]
		which satisfies $(q',\epsilon) \mathrel{T} (p',\epsilon)$ by definition of $T$.

	\item For transitions produced by rule \eqref{rho mid},
		\[
			(q,w)\xtransto[\rho(\M)]{a} (q,wa),
		\]
		we have $(q,wa)\in Q^{\rho(\M)}$. Obviously, $wa\neq\epsilon$, and so by
		the definition of $Q^{\rho(\M)}$, there must be at least one transition
		$q\xtransto[\M]{b} q'$ with $wa\lneqq \R(b)$. By the definition of $T$, we
		have $(q,p)\in S$ and so the simulation $S$ provides us with some
		$p\xtransto[\N]{b} p'$ in $\N$ with $(q',p')\in S$.
		Hence, $(p,wa)\in Q^{\rho(\N)}$ and so we have $(p,w)\xtransto[\rho(\N)]{a}
		(p,wa)$ by rule \eqref{rho mid} and $(q,wa)\mathrel{T} (p,wa)$ by definition of $T$.
		\qedhere
	\end{enumerate}
\end{proofappendix}

\begin{proofappendix}{rho deterministic}
	Consider $(q,w)\in Q^{\rho(\M)}$ and transitions
	\[
	(q,w)
	\xtransto{a} (q_1,w_1)
	\qquad\text{and}\qquad
	(q,w)
	\xtransto{a} (q_2,w_2)
	\qquad\text{in }\rho(\M).
	\]
	We show $(q_1,w_1) = (q_2,w_2)$ by case distinction:
	\begin{enumerate}
		\item Case: there is $b\in \dom(\R)$ with $\R(b) = wa$.
		Then there is a unique such $b$ because $\R$ is prefix-free. Also because
		$\R$ is prefix-free, there is no $b'\in \dom(\R)$ such that $wa\lneqq
		\R(b')$, and so $(q,wa)\notin Q^{\rho(\M)}$ and so neither of the transitions
		were produced by rule \eqref{rho mid}.
		Thus, both transitions come from rule \eqref{rho end} and thus we have $w_1
		= \epsilon = w_2$ and transitions:
		\[
		q\xtransto[\M]{b} q_1
		\qquad\text{and}\qquad q\xtransto[\M]{b} q_2.
		\]
		By assumption, $\M$ is deterministic, so $q_1=q_2$, and so $(q_1,w_1) = (q_2,w_2)$.
		
		\item Case: there is no $b\in \dom(\R)$ with $\R(b) = wa$. Thus, neither of the transitions can be produced by rule \eqref{rho end} and so both were produced by rule \eqref{rho mid}. Then, we necessarily have
		\[
		(q_1,w_1) = (q,wa) = (q_2,w_2).
		\tag*{\qedhere}
		\]
	\end{enumerate}
\end{proofappendix}

\begin{lemma}
	\label{rho basic}
	For all $\M\in \LTS(B)$ and $b\in \dom(\R)$ of an action code $\R\in\Code(A,B)$,
	and $q\in Q^{\M}$:
	\[
	q\xtransto{b} q'
	\text{ in }\M
	\qquad\text{iff}\qquad
	(q,\epsilon)\xTransto{\R(b)} (q',\epsilon)
	\text{ in }\rho(\M).
	\]
\end{lemma}

\begin{proofappendix}{rho basic}
	For $(\Rightarrow)$,
	$\R$ is defined for $b$ and we yield $\R(b) = a_1\cdots a_n$
	with $1 \le n$ and $a_i \in A$ for all $1\le i \le n$.
	Given $q\xtransto{b} q'$ in $\M$,
	we can construct a path in $\rho_\R(\M)$
	by $n-1$ applications of rule \eqref{rho mid} and one application of rule \eqref{rho end}:
	\[
	(q, \epsilon) \xtransto{a_1} (q, a_1)
	\xtransto{a_2} (q, a_1a_2)
	\cdots \xtransto{a_{n-1}} (q, a_1\cdots a_{n_1})
	\xtransto{a_n} (q', \epsilon)
	\qquad\text{ in }\rho_\R(\M),
	\]
	with other notation for $\R(b) = a_1\cdots a_n$:
	\[
	(q, \epsilon) \xTransto{\R(b)} (q', \epsilon)
	\qquad\text{ in }\rho_\R(\M),
	\]
	
	For $(\Leftarrow)$, consider a generalized transition $(q,\epsilon) \xTransto{\R(b)}
	(q',\epsilon)$ in $\rho(\M)$. Of course $\R(b) \in A^+$ so we can write it as
	$\R(b) = wa$ for $w\in A^*$ and $a\in A$.
	Then, we can consider the intermediate state of the given run for $\R(b)$:
	\[
	(q,\epsilon) \xTransto[\rho(\M)]{w} (\bar q, \bar w) \xtransto[\rho(\M)]{a}
	(q',\epsilon).
	\]
	All prefixes $v \le w$ satisfy $v\lneqq \R(b)$, so the first transitions for
	$w$ must have been produced by rule \eqref{rho mid}. Hence, $(\bar q,\bar w) = (q,w)$.
	The final $a$-transition must have been produced by \eqref{rho end} because the second component of the tuple is $\epsilon$. The
	assumption of this rule contains $q\xtransto[\M]{b} q'$, as desired.
\end{proofappendix}

\begin{lemma}
	\label{rho transition}
	For every transition $(q,w)\xtransto{a} (q',w')$ in $\rho(\M)$, there is some transition $q\xtransto{b} q''$ in $\M$ and $u\in A^*$ such that $wau = \R(b)$
	and $(q',w') \xTransto{u} (q'',\epsilon)$ in $\rho_\R(\M)$.
\end{lemma}

\begin{proofappendix}{rho transition}
	For the transition $(q,w)\xtransto{a} (q',w')$, distinguish two cases:
	\begin{itemize}
		\item If $w'=\epsilon$, then the transition must have been produced by rule
		\eqref{rho end}. Thus, the rule assumption contains a transition
		$q\xtransto[\M]{b} q'$ with $wa = \R(b)$. This is the desired witness
		$q'' := q'$ and $u=\epsilon$.
		
		\item If $w'\neq\epsilon$, then the transition must have been produced by
		rule \eqref{rho mid} and so $q'=q$ and $w' = wa$.
		The definition of $Q^{\rho(\M)}$ unfolded for $(q,wa)\in Q^{\rho(\M)}$
		yields that there exists a transition $q\xtransto[\M]{b} q''$ with $wa\lneqq
		\R(b)$. Let $v\in A^*$ and $b\in A$ such that $wavb= \R(b)$.
		Since $wav\lneqq \R(b)$, we can produce further transitions
		using rule \eqref{rho mid} for the letters in $v\in A^*$ and can conclude
		using rule \eqref{rho end} for $b$:
		\[
		(q',w') = (q,wa) \xTransto[\rho(\M)]{~~v~~} (q,wau)
		\xtransto[\rho(\M)]{~b~} (q'',\epsilon).
		\]
		This is the desired transition (with $u:= vb$).
		\qedhere
	\end{itemize}
\end{proofappendix}

\begin{proofappendix}{galois_connection refinement}
	Fix systems $\N\in \LTS(B)$, $\M\in \LTS(A)$, and as usual we
	omit index $\R$ from $\alpha$ and $\rho$.
	\begin{enumerate}
		\item For direction $(\Rightarrow)$, we assume that
		$\N$ restricts to the subclass $\LTS(\dom(\R)) \subseteq \LTS(B)$.
		Hence, in the following we simply put $B:=\dom(\R)$ without loss of
		generality.
		Consider $\rho(\N) \sqsubseteq \M$,
		witnessed by the simulation $S\subseteq Q^{\rho(\N)}\times Q^{\M}$.
		We verify that we have a simulation $T$ between $\N$ and $\alpha(\M)$ defined by:
		\begin{eqnarray*}
			T := \set{ (p,q) \in Q^{\N}\times Q^{\alpha(\M)} \mid 
				\big((p,\epsilon), q\big) \in S}
		\end{eqnarray*}
		The definition of $T$ is well-typed because $Q^{\alpha(\M)}\subseteq Q^{\M}$, and moreover,
		\[
			\set{(p,\epsilon)\mid p\in Q^{\N}} \subseteq Q^{\rho(\N)}.
		\]
		The relation $T$ relates the initial states $q_0^\N \mathrel{T} q_0^{\alpha(\M)}$ because
		\[
			(q_0^\N,\epsilon)
			= q_0^{\rho(\N)} \mathrel{S} q_0^\M
			= q_0^{\alpha(\M)}.
		\]
		Now, suppose $p \; T \; q$ and $p \xtransto[\N]{b} p'$.
		By assumption $\dom(\R) = B$, we can apply \autoref{rho basic} and obtain
		\[
			(p,\epsilon)\xTransto[\rho(\N)]{\R(b)} (p',\epsilon)
			\qquad\text{ in }\rho(\N).
		\]
		The simulation $S$ from $\rho(\N)$ to $\M$ transforms this into a path
		\[
			q \xTransto[\M]{\R(b)} q'\text{ in }\M
			\quad\text{ with }\quad
			(p',\epsilon)\mathrel{S} q'.
		\]
		Since $q \in Q^{\alpha(\M)}$ also $q' \in Q^{\alpha(\M)}$ and thus
		\[
			q \xtransto[\alpha(\M)]{b} q'\text{ in }\alpha(\M)
		\]
		and moreover $p' \mathrel{T} q'$.

		\item For direction $(\Leftarrow)$, assume $\N \sqsubseteq \alpha(\M)$.
		Let
		$S\subseteq Q^{\N}\times Q^{\alpha(\M)}$ be a simulation relation
		from $\N$ to $\alpha(\M)$.  We define the relations $\bar S
		\subseteq Q^{\rho(N)}\times Q^{\alpha(\M)}$
		and
		$T\subseteq Q^{\rho(\N)}\times Q^\M$:
		\[
			(p,\epsilon)\mathrel{\bar S} q
			\quad:\Leftrightarrow\quad
			p\mathrel{S} q
		\]
		\begin{eqnarray*}
			(p, w) \mathrel{T} q & :\Leftrightarrow &
			\exists  \bar q\in Q^{\alpha(\M)}\colon  p \mathrel{S} \bar q  
			\wedge \bar q \xTransto[\M]{w} q
		\end{eqnarray*}
		So visually, every related pair $(p,w)\mathrel{T} q$ entails
		states of the following form:
		\[
			\begin{tikzpicture}[lts,x=3.5cm,y=-2.5cm]
			\node[state] (pr0) {$(p,\epsilon)$};
			\node[state] (q') at (1,0) {$\bar q$};
			\node[state] (pr) at (0,1) {$(p,w)$};
			\node[state] (q) at (1,1) {$q$};
			\drawRealm{\rho(\N)}{(pr0) (pr)}
			\drawRealm{\M}{(q') (q)}
			\path[generalized transition] (pr0) to node[swap] {$w$} (pr);
			\path[generalized transition] (q') to node {$w$} (q);
			\draw (pr0) -- node[anchor=center,fill=white] {{$\bar S$}} (q');
			\draw (pr) -- node[anchor=center,fill=white] {$T$} (q);
			\end{tikzpicture}
		\]
	We verify that $T$ is a simulation from $\rho(\N)$ to $\M$.

	Picking $\bar q := q$ and $w:=\epsilon$ shows that the related initial states $q_0^\N \mathrel{S} q_0^{\alpha(\M)}$ of $\N$ and $\alpha(\M)$ imply
	\[
		q_0^{\rho(\N)} = (q_0^{\N},\epsilon) \mathrel{T} q_0^{\M}.
	\]

	Suppose $(p, w) \mathrel{T} q$ and $(p, w) \xtransto{a}_{\rho(\N)} (p', w')$.
	\autoref{rho transition} shows that $wa\in A^+$ sits \textqt{below} some
	$b\in B$ in the action code $\R$: concretely, there are $u\in A^*$, and $b\in B$ such that:
		\[
			(p',w')\xTransto[\rho(\N)]{u}(p'',\epsilon)
			\qquad\text{and}\qquad
			wau = \R(b)
			\qquad\text{and}\qquad
			p\xtransto[\N]{b}p''.
		\]
		We have the solid (i.e.\ non-dotted) part of the following picture:
		\[
			\begin{tikzpicture}[lts,x=3.0cm,y=-1.8cm]
			\node[state] (pr0) {$(p,\epsilon)$};
			\node[state] (pr) at (0,1) {$(p,w)$};
			\node[state] (p'r') at (0,2) {$(p',w')$};
			\node[state] (p''r0) at (0,3) {$(p'',\epsilon)$};
			\path[generalized transition] (pr0) to node[swap] {$w$} (pr);
			\path[transition] (pr) to node[swap] {$a$} (p'r');
			\path[generalized transition] (p'r') to node[swap] {$u$} (p''r0);
			\drawRealm{\rho(\N)}{(pr0) (p''r0)}
			\begin{scope}[xshift=3cm]
				\node[state] (bar q) at (0,0) {$\bar q$};
				\node[state] (q) at (0,1) {$q$};
				\node[state] (q') at (0,2) {$q'$};
				\node[state] (q'') at (0,3) {$q''$};
				\path[generalized transition] (bar q) to node {$w$} (q);
				\path[transition,dotted] (q) to node {$a$} (q');
				\path[generalized transition,dotted] (q') to node {$u$} (q'');
				\drawRealm{\M}{(bar q) (q'')}
			\end{scope}
			\draw (pr0) edge node[anchor=center,fill=white] {{$\bar S$}} (bar q);
			\draw (pr) edge node[anchor=center,fill=white] {$T$} (q);
			\draw (p''r0) edge[dotted] node[anchor=center,fill=white] {$\bar S$} (q'');

			\begin{scope}[xshift=6cm]
				\node[state] (p) at (0,0) {$p$};
				\node[state] (p') at (0,3) {$p''$};
				\path[transition] (p) to node[swap] {$b$} (p');
				\drawRealm{\N}{(p) (p')}
				\node[state] (bar q alpha) at (1,0) {$\bar q$};
				\node[state] (q'' alpha) at (p' -| bar q alpha) {$q''$};
				\draw[transition,dotted] (bar q alpha) edge node {$b$} (q'' alpha);
				\drawRealm{\alpha(\M)}{(bar q alpha) (q'' alpha)}
				\draw[dotted] (p') -- node[anchor=center,fill=white] {$S$} (q'' alpha);
				\draw (p) -- node[anchor=center,fill=white] {$S$} (bar q alpha);
			\end{scope}
			\end{tikzpicture}
		\]

	Using the simulation property of $p\mathrel{S} \bar q$, we obtain $q''\in
	Q^{\alpha(\M)}$ with $\bar q\xtransto[\alpha(\M)]{b} q''$ and $p''\mathrel{S} q''$.
	The transition $\bar q\xtransto{b} q''$ in $\alpha(\M)$ must have come from a path $\bar q\xTransto{\R(b)} q''$ in $\M$. Denote the intermediate states for the decomposition $\R(b) = wau$ by $q_w, q'\in Q^\M$:
	\[
		\bar q \xTransto[\M]{w} q_w \xtransto[\M]{a} q'\xTransto{u} q'' 
	\]
	The assumption that $\M$ is deterministic enforces that $q= q_w$, so $q\xTransto[\M]{au} q''$ as shown in the picture. Also, $p''\mathrel{S} q''$ directly shows $(p'',\epsilon)\mathrel{\bar S} q''$. Now, the picture is complete.
	For the final proof that $T$ relates $(p',w')$ and $q'$, we
	distinguish cases:
	\begin{enumerate}
		\item Case $u= \epsilon$:
			Then, $p'=p''$, $w'=\epsilon$ and $q' =q''$. Thus, $\big((p',w'),q'\big)\in \bar S\subseteq T$.
		\item Case $u \neq \epsilon$:
			Then, $wa\lneqq \R(b)$ and so $wa\neq \R(b')$ for all $b'\in B$ by prefix-freeness.
			So the $a$-transition can only come from rule \eqref{rho mid}; hence $p=p'$ and $w'=wa$.
			Finally, $(p',w') = (p,wa) \mathrel{T} q'$ by the definition of $T$ and $\bar q\xTransto{wa} q'$.
		\qedhere
	\end{enumerate}
	\end{enumerate}
\end{proofappendix}

\begin{iflong}
\begin{proofappendix}[Details for]{why functional}%
	\twnote{}%
	\fvnote{}%
	Contraction $\alpha$ has no left-adjoint if we allow the action codes
	to relate multiple concrete words $w_1,w_2\in A^+$ with the same abstract
	symbol $b\in B$.
	We show the special case for both words having length 1, which generalizes to general words.
	Assume there are two symbols $a_1,a_2\in A,a_1\neq a_2$ both related to $b\in B$. Denote a system with two
	states and one transition directly by $(\bullet \xtransto{b}
	\bullet)$ (the initial state is the left-hand one). Then, we have
	\[
		\alpha(\bullet \xtransto{a_1} \bullet)
		= (\bullet \xtransto{b} \bullet)
		= \alpha(\bullet \xtransto{a_2} \bullet).
	\]
	Then, there exists no left adjoint $\rho$. Such an adjoint, applied to $(\bullet \xtransto{b}
	\bullet)$ would need to satisfy
	\[
		\rho(\bullet \xtransto{b} \bullet)
		\sqsubseteq (\bullet \xtransto{a_1} \bullet)
		\quad\text{and}\quad
		\rho(\bullet \xtransto{b} \bullet)
		\sqsubseteq (\bullet \xtransto{a_2} \bullet)
	\]
	Hence, the initial state in $\rho(\bullet \xtransto{b}\bullet)$ is a
	leaf state, i.e.~$\rho(\bullet \xtransto{b}\bullet)$ is the bottom
	element for the simulation order $\sqsubseteq$. This easily leads to a
	contradiction after using the Galois connection in the other direction again.
\end{proofappendix}
\end{iflong}

\begin{lemma}\label{gammatrans} \twnote{}
	For every $\M\in \LTS(B)$, $q\in Q^\M$, $\R \in \Code(A,B)$, $\SI$ being the identity relation on $A$, and $b\in \dom(\R)$:
	\begin{enumerate}
	\item Whenever $q\xtransto[\M]{b} q'$
	then $(q,\epsilon)\xTransto[\gamma(\M)]{\R(b)} (q',\epsilon)$.
	\item Whenever $(q,\epsilon)\xTransto[\gamma(\M)]{\R(b)} \bar q$,
	then $\bar q = (q',\epsilon)$ for some $q'\in Q^\M$ with $q\xtransto[\M]{b} q'$.
	\end{enumerate}
\end{lemma}

\begin{proofappendix}{gammatrans}
	\begin{enumerate}
	\item Given $q\xtransto[\M]{b} q'$ in $\M$, write $\R(b)\in A^+$ as
	$\R(b) = w a$ for $w\in A^*$ and $a\in A$.
	Write $w = w_1\cdots w_n$, with $n\in \Nat$ and $w_i\in A$, for $1\le i \le n$.
	For every $1\le i\le n$, the word $w_1\cdots w_i\in A^*$ is contained in $N$
	because $w_1\cdots w_i\lneqq \R(b)$. Hence, rule \eqref{rule CW1} provides us with transitions
	\[
		(q,\epsilon)
		\xtransto[\gamma(\M)]{w_1}
		(q,w_1)
		\xtransto[\gamma(\M)]{w_2}
		(q,w_1w_2)
		\cdots
		\xtransto[\gamma(\M)]{}
		(q,w_1\cdots w_n)
		= (q,w)
	\]
	Finally, rule \eqref{rule CW2} has the assumptions $q\xtransto[\M]{b} q'$ and
	$wa = \R(b)$ fulfilled, so we have
	\[
		(q,\epsilon) \xTransto[\gamma(\M)]{w}
		(q,w) \xtransto[\gamma(\M)]{a}
		(q',\epsilon),
	\]
	i.e.~$(q,\epsilon) \xTransto[\gamma(\M)]{\R(b)} (q',\epsilon)$, as desired.

	\item Assume $(q,\epsilon)\xTransto[\gamma(\M)]{\R(b)} \bar q$. As before, decompose
	$\R(b)$ into $\R(b) = wa$ for $w\in A^*$ and $a\in A$. Write $w=w_1\cdots
	w_n$ for $w_i\in A$, $n\in \Nat$. For every $i$, $1\le i < n$, any transition
	\[
		(q,w_1\cdots w_i)
		\xtransto[\gamma(\M)]{w_{i+1}}
		p
	\]
	can only be produced by rule \eqref{rule CW1}, because $w_1\cdots w_iw_{i+1} \in N$.
	Hence, $p = (q,w_1\cdots w_{i+})$. Thus, the given run for $\R(b)\in A^+$ is of the form
	\[
		(q, \epsilon)
		\xtransto[\gamma(\M)]{w_{1}}
		(q, w_1)
		\xtransto[\gamma(\M)]{w_{2}}
		\cdots
		\xtransto[\gamma(\M)]{w_{n}}
		(q, w_1\cdots w_n)
		= (q,w)
		\xtransto[\gamma(\M)]{a}
		\bar q.
	\]
	In order to see that the final $a$-transition is produced by rule \eqref{rule
	CW2}, note that $wa = \R(b)$ implies that $wa\notin N$ by prefix-freeness
	\eqref{prefix-free}, and so rule \eqref{rule CW1} can not have produced this
	$a$-transition to $\bar q$. Obviously, rule \eqref{rule CW3} does not match
	either, and so only rule \eqref{rule CW2} is left.
	Thus, there is some $q\xtransto[\M]{b} q'$ and $\bar q$ is of the form $\bar q = (q',\epsilon)$, as desired.
	\qedhere
	\end{enumerate}
\end{proofappendix}

\begin{proofappendix}{gamma galois}
	In the present proof, we restrict to the special case where $\SI\subseteq A\times A$ is the identity relation on $A$, which we thus omit from the index in $\gamma$. The proof for a general $\SI$ is entirely formalized in Coq.
	Since we have only one action code $\R$ at hand, we omit the index $\R$ for
	$\alpha$ and $\gamma$. We prove both directions seperately:
	\begin{description}
	\item[$(\Rightarrow)$]
		Let $\alpha(\M)\sqsubseteq \N$ be witnessed by the simulation $S\sqsubseteq
		Q^{\alpha(\M)}\times Q^{\N}$. We show that
		\(
			T\sqsubseteq Q^\M \times Q^{\gamma(\N)}
		\)
		defined by
		\[
			T :=
				\begin{array}[t]{l}
				\set{(p', (q,w)) \mid p'\in Q^\M, (q,w)\in Q^{\gamma(\N)}, \exists p\in Q^\M, p\xTransto[\M]{w} p', (p,q)\in S }
				\\
				\cup\,\set{(p, \chi) \mid p\in Q^\M}
				\end{array}
		\]
		is a simulation. Note that if $\chi$ is omitted from $\gamma(\N)$ if it is
		not reachable, and then we also omit it from $T$. For the initial states,
		we immediately have
		\[
			(q_0^\M, q_0^{\gamma(\N)})
			= (q_0^\M, (q_0^\N,\epsilon)) \in T
		\]
		by the definition of $T$, because $q_0^\M \xTransto{\epsilon} q_0^\M$ and
		$(q_0^\M, q_0^{\N}) = (q_0^{\alpha(\M)}, q_0^{\N}) \in S$.

		Since $T$ is defined as a
		union, we can verify the two parts seperately:
		\begin{enumerate}
		\item Consider $(p', (q,w)) \in T$ and $p'\xtransto[\M]{a} p''$. By definition of $T$, there is some $p\in Q^\M$ with
		\[
			p \xTransto[\M]{w} p'
			\qquad\text{and}\qquad (p,q) \in S.
		\]
		We distinguish whether $wa \in N$ and $wa\in \Im(\R)$:
		\begin{itemize}
		\item If $wa\in N$, then we have $(q,w)\xtransto[\gamma(\N)]{a} (q,wa)$ by
		rule \eqref{rule CW1} and $(p'',(q,wa))\in T$ by definition of $T$.

		\item If $wa\notin N$ and $wa\in \Im(\R)$, then there is some $b\in \dom(\R)$ with
		$\R(b) = wa$. We have $p\xTransto[\M]{w} p' \xtransto[\M]{a} p''$ and so
		$p\xtransto[\alpha(\M)]{b} p''$ by the definition of $\alpha(\M)$.
		Using that $S$ is a simulation, we obtain a transition $q\xtransto[\N]{b}
		q'$ in $\N$. By rule \eqref{rule CW2}, this translates into a transition
		$(q,w)\xtransto[\gamma(\N)]{a} (q',\epsilon)$ in $\gamma(\N)$.
		By definition of $T$, we find $(p'', (q',\epsilon)) \in T$.

		\item If $wa\notin N$ and $wa\notin \Im(\R)$, then we have
			$(q,w)\xtransto[\gamma(\N)] \chi$ by rule \eqref{rule CW3}. In this case,
			Then, we also have $(p',\chi)\in T$.
		\end{itemize}

		\item Consider $(p, \chi)\in T$ and $p\xtransto[\M]{a} p'$ in $\M$. We have
			$\chi \xtransto[\gamma(\N)]{a} \chi$ by rule \eqref{rule CW4} and
			$(p',\chi)\in T$, again.
		\end{enumerate}
	\item[$(\Leftarrow)$]
		Assume $M\sqsubseteq \gamma_\R(\N)$ in $\LTS(A)$, witnessed by a simulation
		$S\subseteq Q^\M\times Q^{\gamma(\N)}$. Define $T\subseteq Q^{\alpha(\M)}\times Q^\N$ by
		\[
			T := \set{
				(p,q) \in Q^{\alpha(\M)}\times Q^\N
				\mid 
				(p, (q,\epsilon)) \in S
			}.
		\]
		Here, we use that $Q^{\alpha(\M)} \subseteq Q^\M$. For the initial states,
		note that $(q_0^{\alpha(\M)}, q_0^\N) \in T$ because
		\[
			(q_0^{\M}, (q_0^\N,\epsilon))
			= (q_0^{\M}, q_0^{\gamma(\N)}) \in S.
		\]
		For the remaining verification,
		consider $(p,q)\in T$ and a transition $p\xtransto[\alpha(\M)]{b} p'$ in
		$\alpha(\M)$. By the definition of $\alpha$, we have $b\in \dom(\R)$ and a
		run
		\[
			p\xTransto[\M]{\R(b)} p'
			\qquad\text{in }\M.
		\]
		Using that $S$ is a simulation and that $(p,(q,\epsilon))\in S$, this yields a run
		\[
			(q,\epsilon) \xTransto[\gamma(\N)]{\R(b)} q'
			\qquad\text{in }\gamma(\N)
			\qquad\text{with}\qquad
			(p',q') \in S.
		\]
		We do not know yet in which of the two components of $Q^{\gamma(\N)} =
		(Q^\N\times N) \cup \set{chi}$ the state $q'$ is. We investigate by
		decomposing the run for $\R(b)\in A^+$ into $wa=\R(b)$ for $w\in A^*$ and
		$a\in A$, calling the intermediate state $\bar q\in Q^{\gamma(\N)}$:
		\[
			(q,\epsilon) \xTransto[\gamma(\N)]{w}
			\bar q \xtransto[\gamma(\N)]{a}
			q'
			\qquad\text{in }\gamma(\N).
		\]
		Since $w\lneqq \R(b)$, we have $w\in N$. Looking at the rules
		for the transitions of $\gamma(\N)$, we see that the only option for the transitions in $(q,\epsilon) \xTransto[\gamma(\N)]{w} \bar q$ with $w\in N$ is via rule \eqref{rule CW1}, so we necessarily obtain $\bar q = (q,w)$:
		\[
			(q,\epsilon) \xTransto[\gamma(\N)]{w}
			(q,w) \xtransto[\gamma(\N)]{a}
			q'
		\]
		Using that $wa = \R(b)$, only rule \eqref{rule CW2} can have produced the
		transition $(q,w) \xtransto{a} q'$, hence $q' = (q'',\epsilon)$ for some
		$q''\in Q^\N$ with $q\xtransto[\N]{b} q''$ in $\N$.
		In total, we have $(p',(q'',\epsilon)) = (p', q') \in S$ and by the
		definition of $T$:
		\[
			(p',q'')\in T
			\qquad
			q \xtransto[\N]{b} q''
			\quad\text{in $\N$}.
		\]
		This shows that $T$ is indeed a simulation.
		\qedhere
	\end{description}
\end{proofappendix}

\begin{proofappendix}{gamma monotone}
	Consider $\M\sqsubseteq \N$ in $\LTS(B)$. By the reflexivity, we have
	\[
		\gamma_{\R,\SI}(\M) \sqsubseteq \gamma_{\R,\SI}(\M)
		\qquad\text{in }\LTS(A).
	\]
	Applying the Galois connection (\autoref{gamma galois}) from right to left yields
	\[
		\alpha_\R(\gamma_{\R,\SI}(\M)) \sqsubseteq \M
		\qquad\text{in }\LTS(B).
	\]
	By transitivity of $\sqsubseteq$ and $\M\sqsubseteq \N$, we obtain
	\[
		\alpha_\R(\gamma_{\R,\SI}(\M)) \sqsubseteq \N
		\qquad\text{in }\LTS(B).
	\]
	Applying the Galois connection (\autoref{gamma galois}) conversely from left
	to right yields the desired
	\[
		\gamma_{\R,\SI}(\M) \sqsubseteq \gamma_{\R,\SI}(\N)
		\qquad\text{in }\LTS(A).
		\tag*{\qedhere}
	\]
\end{proofappendix}

\takeout{}

\begin{proofappendix}{gamma deterministic}
	Again, we restrict to the case where $\SI$ is the identity relation on $A$; the general proof is formalized in Coq.
	We verify the determinacy seperately for the disjoint components of $Q^{\gamma(\M)} :=
	(Q^{\M}\times N) \cup \set{\chi}$
	\begin{enumerate}
	\item For $(q,w)\in Q^{\M}\times N$ and two transitions
	\[
		(q,w) \xtransto[\gamma(\M)]{a} \bar q_1
		\quad
		(q,w) \xtransto[\gamma(\M)]{a} \bar q_2
	\]
	we distinguish cases like in the assumptions of the rules for $\xtransto[\gamma(\M)]{}$:
	\begin{itemize}
	\item If $wa\in N$, then both transitions have been produced by rule \eqref{rule CW1} and so
	$\bar q_1 = (q,wa) = \bar q_2$.
	\item If $wa\notin N$ and $wa\in \Im(\R)$, then there are $b_1,b_2\in \dom(\R)$ with
	\[
		q\xtransto[\M]{b_1} q_1'
		\qquad
		\bar q_1 = (q_1',\epsilon)
		\qquad
		\R(b_1) = wa
	\]
	\[
		q\xtransto[\M]{b_2} q_2'
		\qquad
		\bar q_2 = (q_2',\epsilon)
		\qquad
		\R(b_2) = wa
	\]
	Since $\R\colon B\partialto A^+$ is prefix-free \eqref{prefix-free}, it is in
	particular injective and so $b_1=b_2$. The LTS $\M$ was assumed to be
	deterministic, thus $q_1' =q_2'$ and so $\bar q_1 = (q_1',\epsilon) = (q_2',
	\epsilon) = \bar q_2$.

	\item If $wa\notin N$ and $wa\notin \Im(\R)$, then both transitions have been
	produced by rule \eqref{rule CW3} and so $\bar q_1 = \chi = \bar q_2$.
	\end{itemize}

	\item Any two outgoing transitions of $\chi$
	\[
		\chi \xtransto[\gamma(\M)]{a} \bar q_1
		\quad\text{and}\quad
		\chi \xtransto[\gamma(\M)]{a} \bar q_2
	\]
	have necessarily been created by \eqref{rule CW4}, and so $\bar q_1 = \chi=\bar q_2$.
	\qedhere
	\end{enumerate}
\end{proofappendix}

\begin{proofappendix}{gamma galois insertion}
	Again, we restrict to the case where $\SI$ is the identity relation on $A$; the general proof is formalized in Coq.
	Since we have only one action code $\R$ at hand, we omit the index $\R$ in
	$\alpha$ and $\gamma$ in this proof.
	The LTS $\alpha(\gamma(\M))$ has precisely the states
	\[
		Q^{\alpha(\gamma(\M))}
		= \set{(q,\epsilon) \mid q\in Q^{\M}}.
	\]
	In order to see that, we find:
	\begin{description}
	\item[$(\subseteq)$]
		If $\bar q\in Q^{\alpha(\gamma(\M))}$, then there is a word
		$b_1\cdots b_n\in \dom(\R)^*$ and are states $\bar q_1,\ldots, \bar q_n$ with
		\[
			q_0^{\gamma(\M)}
			= (q_0^\M, \epsilon)
			\xtransto[\alpha(\gamma(\M))]{b_1}
			\bar q_1
			\xtransto[\alpha(\gamma(\M))]{b_2}
			\cdots
			\xtransto[\alpha(\gamma(\M))]{b_n}
			\bar q_n = \bar q.
			\qquad\text{in }\alpha(\gamma(\M)).
		\]
		By definition of $\alpha$, this corresponds to transitions
		\[
			q_0^{\gamma(\M)}
			= (q_0^\M, \epsilon)
			\xTransto[\gamma(\M)]{\R(b_1)}
			\bar q_1
			\xTransto[\gamma(\M)]{\R(b_2)}
			\cdots
			\xTransto[\gamma(\M)]{\R(b_n)}
			\bar q_n = \bar q.
			\qquad\text{in }\gamma(\M).
		\]
		Applying \autoref{gammatrans} to every $\bar q_i$, we obtain that $\bar q = (q,\epsilon)$ for some $q\in Q^\M$.
		Note in particular, $\bar q\neq \chi$ and so $\chi\notin Q^{\alpha(\gamma(\M))}$.
	\item[$(\supseteq)$]
		The converse inclusion iterates the other direction of \autoref{gammatrans}:
		we assume $\M\in \LTS(B)$ to be reachable, hence every state $q\in Q^{\M}$
		is reachable via some word $b_1\cdots b_n \in B^*$ by iterating \autoref{gammatrans}:
		\[
			q_0^\M \xTransto[\M]{b_1\cdots b_n} q.
		\]
		The assumption that $\M\in \LTS(\dom(\R))$ implies that $\R\colon
		B\partialto A^+$ is defined for every $b_i$, and thus $(q,\epsilon)$ is
		reachable in $Q^{\gamma(\M)}$:
		\[
			q_0^{\gamma(\M)} = (q_0^\M,\epsilon) \xTransto[\gamma(\M)]{\R(b_1)\cdots \R(b_n)} (q,\epsilon).
		\]
		And hence, the definition of $\alpha$ then sends this run to
		\[
			q_0^{\alpha(\gamma(\M))} =
			q_0^{\gamma(\M)} = (q_0^\M,\epsilon) \xTransto[\alpha(\gamma(\M))]{b_1\cdots b_n} (q,\epsilon).
		\]
		and so $(q,\epsilon) \in Q^{\alpha(\gamma(\M))}$.
	\end{description}
	The witnessing bijective bisimulation is
	\[
		\phi\colon Q^\M\longrightarrow Q^{\alpha(\gamma(\M))}
		\qquad
		\phi(q) = (q,\epsilon) ~~\in Q^{\alpha(\gamma(\M))} \subseteq Q^{\gamma(\M)}.
	\]
	By our above characterization of $Q^{\alpha(\gamma(\M))}$, $\phi$ is a bijection.
	It remains to verify that $\phi$ is a bisimulation:
	\begin{itemize}
	\item For every transition in $\alpha(\gamma(\M))$, concretely $(q,\epsilon)\xtransto{b} (q',\epsilon)$, we have
		$(q,\epsilon) \xTransto{\R(b)} (q',\epsilon)$ in $\gamma(\M)$ by the definition of $\alpha$.
		By \autoref{gammatrans}, this implies $q\xtransto{b} q'$ in $\M$; and indeed $\phi(q) = (q,\epsilon)$ and $\phi(q') = (q',\epsilon)$.
	\item Conversely, for every transition $q\xtransto{b} q'$ in $\M$, we have a transition
		\[
			(q,\epsilon) \xTransto{\R(b)} (q',\epsilon)
			\qquad
			\text{in }\gamma(\M)
		\]
		by \autoref{gammatrans} and by $b\in \dom(\R)$ provided by the assumption $\M\in \LTS(\dom(\R))$.
		By the definition of $\alpha$, we thus have
		\[
			\phi(q) = (q,\epsilon) \xtransto{b} (q',\epsilon) = \phi(q')
			\qquad
			\text{in }\alpha(\gamma(\M)).
		\]
	\end{itemize}
	In total, $\phi$ is an isomorphism in $\LTS(B)$.
\end{proofappendix}

\begin{proofappendix}[Details of]{remark kleisli}
	Kleisli composition is a recipe to compose maps of the form $C\to T(B)$ and
	$B\to T(A)$ to a map of type $C\to T(A)$, where $T$ is a monad. In our case, the
	monad is $T(X) = X^+ + 1$ where $1$ is an arbitrary singleton and $+$ denotes disjoint union. This monad $T$ itself is a combination of two monads:
	$S(X) = X^+$ is the free semigroup-monad. Monads corresponds to
	algebraic theories and the algebraic theory corresponding to $S$ is that of
	semigroups. The monad $P(X) = X + 1$ is called the \emph{maybe monad} (or
	sometimes called \emph{optional} in programming), which allows to model
	partial maps. The algebraic theory corresponding to $P$ is that of
	\emph{pointed sets} (the theory consists of one nullary operation).
	Warning: even though $T(X) = P(S(X)) = X^+ + 1$ and $M(X) = X^*$ are
	naturally isomorphic functors, they are different monads, because $M$ is the
	\emph{list monad}, whose corresponding algebraic theory is that of monoids.
\end{proofappendix}

\begin{proofappendix}{lem:composition}
	We need to show that $(\R * \S) \colon C\partialto A^+$ is prefix-free. To
	this end, consider $c,c'\in \dom(\R*\S)$ with
	\[
		(\R * \S)(c) \le (\R * \S)(c').
	\]
	Since $\R*\S$ is defined for both $c$ and $c'$ we can spell out the words as
	\begin{align*}
		(\R * \S)(c) &= \R(b_1)\cdots \R(b_n) & \text{for }n\in \Nat
			\text{ and }\S(c) &= b_1\cdots b_n\\
		(\R * \S)(c') &= \R(b_1')\cdots \R(b_m') & \text{for }m\in \Nat
			\text{ and }\S(c') &= b_1'\cdots b_m'.\\
	\end{align*}
	Note that we do not know yet whether $n$ or $m$ is bigger! We only know that
	\begin{equation}
		\R(b_1)\cdots \R(b_n)
		\le \R(b_1')\cdots \R(b_m').
		\label{RbPrefixOf}
	\end{equation}
	We now show by induction that
	\[
		\text{for all }i\text{ with } 0\le i\le \min(n,m)\colon
		~~
		b_i = b_i'.
	\]
	\begin{itemize}
	\item In the base case $i=0$, there is nothing to be shown.
	\item In the step for $i$, assume that we have 
	$\forall 0\le j < i\colon b_i = b_i'$ as the induction hypothesis. Thus, we also have
	$\R(b_j) = \R(b_j')$ for all $j< i$ and so the words
	\[
		\R(b_1)\cdots \R(b_i) \cdots R(b_n)
		\text{ and } \R(b_1')\cdots \R(b_i') \cdots  \R(b_m').
	\]
	have a common prefix $\R(b_1)\cdots \R(b_{i-1}) =
	\R(b_1')\cdots \R(b_{i-1}')$. For general $u,v,w\in C^*$, if $uv \le uw$, then $v\le
	w$. So after removing the common prefix from both sides of \eqref{RbPrefixOf}, obtain
	\[
		\R(b_i)\cdots \R(b_n)
		\le \R(b_i')\cdots \R(b_m').
	\]
	In such a scenario, we either have $\R(b_i) \le \R(b_i')$ or $\R(b_i) \ge
	\R(b_i')$. Since $\R$ is prefix-free \eqref{prefix-free}, we obtain $b_i =
	b_i'$ in either case.
	\end{itemize}
	We can now use the inductively proven statement to show that $b_1\cdots b_n \le b_1'\cdots b_m'$ by case distinction:
	\begin{itemize}
	\item If $\min(n,m) = m$, i.e.~$n \ge m$, then we can remove the common
		prefix $\R(b_1)\cdots \R(b_m) = R(b_1')\cdots \R(b_m')$
		from both sides of \eqref{RbPrefixOf} in order to obtain
		\[
			\R(b_{m+1})\cdots \R(b_n)
			\le \epsilon
		\]
		Since all $\R(b_i)\in A^+$ for all $1\le i \le n$, we necessarily have $m = n$. This implies $b_1\cdots b_n \le b_1'\cdots b_m'$ (both sides are identical).

	\item If $\min(n,m) = n$, i.e.~$n \le m$, then we directly have
		$b_1\cdots b_n \le b_1'\cdots b_m'$.
	\end{itemize}
	So in any case, $\S(c) = b_1\cdots b_n \le b_1'\cdots b_m' = \S(c')$. Using
	that $\S$ is prefix-free \eqref{prefix-free}, we conclude $c=c'$, as desired.
\end{proofappendix}

\begin{proofappendix}[Proof of \autoref{thm code composition}]{alpha code composition}
	We show that the systems
	$\alpha_{\R*\S}(\M)$ and $\alpha_{\S}(\alpha_{\R}(\M))$ are even identical.
	Note that we have state sets:
	\[
		Q^{\alpha_{\R*\S}(\M)} \subseteq Q^\M
		\text{ and }
		Q^{\alpha_{\R}(\alpha_{\S}(\M))}
		\subseteq Q^{\alpha_{\S}(\M)}
		\subseteq Q^\M
	\]
	which are both subsets of $Q^\M$. Their initial states
	are identical, because they are both $q_0^\M$.
	We establish the isomorphism by simultaneously showing that the state sets
	match and that the transitions match:
	\begin{enumerate}
	\item[$(\subseteq)$]
		Consider a transition
		\[
			q \xtransto{c} q'
			\qquad
			\text{in }\alpha_{\R*\S}(\M)
		\]
		for which we already assume that $q\in Q^{\alpha_\R(\alpha_\S(\M))}$.
		Thus, we have
		\[
			q \xTransto{(\R*\S)(c)} q'
			\qquad
			\text{in }\M.
		\]
		By the definition of $\R*\S$, we have $c\in \dom(\S)$ and $b_1,\ldots,b_n\in \dom(\R)$ with $\S(c) = b_1\cdots b_n$, so above sequence can be rewritten as
		\[
			q\xTransto{\R(b_1)\cdots \R(b_n)} q'
			\qquad
			\text{in }\M
		\]
		or equivalently
		\[
			q\xTransto{\R(b_1)}\cdots \xTransto{\R(b_n)} q'
			\qquad
			\text{in }\M
		\]
		By definition of $\alpha_\R$, we have
		\[
			q\xtransto{b_1}\cdots \xtransto{b_n} q'
			\qquad
			\text{in }\alpha_\R(\M)
		\]
		or equivalently
		\[
			q\xTransto{\S(c)} q'
			\qquad
			\text{in }\alpha_\R(\M).
		\]
		Finally, by the definition of $\alpha_\S$, this yields
		\[
			q\xtransto{c} q'
			\qquad
			\text{in }\alpha_\S(\alpha_\R(\M)).
		\]
		This first shows that all states of $\alpha_{\R*\S}(\M)$ are also contained
		in $\alpha_\R(\alpha_\S(\M))$ and secondly that the transitions are
		included, too.
	\item[$(\supseteq)$]
		The converse direction is analogous, starting with a transition
		\[
			q\xtransto{c} q'
			\qquad
			\text{in }\alpha_\S(\alpha_\R(\M))
		\]
		for which we know $q\in Q^{\alpha_{\R*\S}(\M)}$ already. Thus, $c\in
		\dom(\S)$ and we obtain
		\[
			q\xTransto{\S(c)} q'
			\qquad
			\text{in }\alpha_\R(\M).
		\]
		With $b_1\cdots b_n = \S(c)$, we have
		\[
			q\xtransto{b_1}\cdots \xtransto{b_n} q'
			\qquad
			\text{in }\alpha_\R(\M).
		\]
		This implies that $b_i\in \dom(\R)$ for every $1\le i\le n$ and moreover
		\[
			\big(q\xTransto{\R(b_1)}\cdots \xTransto{\R(b_n)} q'\big)
			= \big(q\xTransto{\R(b_1)\cdots \R(b_n)} q'\big)
			\qquad
			\text{in }\M.
		\]
		Since all $b_i\in \dom(\R)$ and $c\in \dom(\S)$, we find that $c\in \dom(\R*\S)$ and so
		$(\R*\S)(c) = \R(b_1)\cdots \R(b_n)$ and
		\[
			q\xTransto{(\R*\S)(c)} q'
			\qquad
			\text{in }\M.
		\]
		Finally, by the definition of $\alpha_{\R*\S}$, we conclude
		\[
			q\xtransto{c} q'
			\qquad
			\text{in }\alpha_{\R*\S}(\M).
			\tag*{\qedhere}
		\]
	\end{enumerate}
\end{proofappendix}

\begin{proofappendix}[Proof of \autoref{thm code composition}]{refinement code composition}
	For the map-based action code $\R\colon B\partialto A^+$, define
	the partial map
	\[
		\R^*\colon B^*\partialto A^*
		\qquad
		\R^*(\epsilon) = \epsilon
		\qquad
		\R^*(b\,w) = \R(b)\,\R^*(w)
		\quad\text{(if both defined)}.
	\]
	By this, we mean that the inductive case $\R^*(b\,w)$ is only defined if both
	$\R(b)$ and $\R^*(w)$ are defined.\footnote{$\R^*$ is also called the Kleisli
	extension of $\R$ for the monad $(-)^*$ on partial maps} With this definition, we have that
	\[
		(\R*\S)(c) = \R^*(\S(c))
		\text{ for all }c\in C.
	\]

	For the isomorphism $h\colon \rho_\R(\rho_\S(\M))\to \rho_{\R*\S}(\M)$, the involved state sets are by \autoref{def: refinement} of the form:
	\[
		Q^{\rho_\S(\M)}
		\subseteq Q^{\M}\times B^*
		\qquad
		Q^{\rho_\R(\rho_\S(\M))}
		\subseteq (Q^{\M}\times B^*)\times A^*
		\qquad
		Q^{\rho_{\R*\S}(\M)}
		\subseteq Q^{\M}\times A^*
	\]
	Define a partial map $h\colon Q^{\rho_\R(\rho_\S(\M))} \to Q^{\rho_{\R*\S}(\M)}$
	by
	\[
		h((q,u),v) = \begin{cases}
			(q,\R^*(u)\,v) &\text{if $\R^*(u)$ is defined}\\
			\text{undefined} &\text{otherwise}.
		\end{cases}
	\]
	Such a partial map is sufficient to establish an isomorphism between the
	reachable parts (cf.~\autoref{def:isomorphism}) of $\rho_\R(\rho_\S(\M))$ and
	$\rho_{\R*\S}(\M)$, because we can show that if $((q,u),v)$ is reachable, then $\R^*(u)$ is defined:
	if $((q,u),v)$ is reachable, then the \emph{shortest} path from the initial state must end
	with a path of the form:
	\[
		((q,\epsilon), \epsilon)
		\xTransto{w}
		((q,u), \epsilon)
		\xTransto{v'}
		((q,u), v)
		\qquad\text{in }\rho_\R(\rho_\S(\M)).
	\]
	If we require that $v'$ is the shortest path from $((q,u),\epsilon)$ to $((q,u),v)$, then all transitions of $v'$ must come from rule \eqref{rho mid}, and so $v=v'$. If we require $w$ to be the shortest path, then by an iterated application of \autoref{rho transition}, we find that $w= \R^*(u)$.

	In order to show that $h$ is a simulation, consider a \emph{reachable} transition
	\[
		((q,u),v) \xtransto{a} ((q',u'),v')
		\qquad\text{in }\rho_\R(\rho_\S(\M)).
	\]
	The transition being reachable implies that $\R^*(u)$ is defined.
	By \autoref{rho transition}, there exists a transition
	\[
		(q,u) \xtransto{b} (q'',u'')
		\qquad\text{in }\rho_\S(\M)
	\]
	with $\R(b) = var$ and some $r\in A^*$ such that
	\[
		((q,u),\epsilon) \xTransto{v}
		((q,u),v) \xtransto{a}
		((q',u'),v') \xTransto{r}
		((q'',u''),\epsilon)
		\qquad\text{in }\rho_\R(\rho_\S(\M)).
	\]
	Applying \autoref{rho transition} to the above $b$-transition in $\rho_\S(\M)$ provides us with
	some $c\in C$, $q'''\in Q^\M$, and $s\in B^*$ with
	$\S(c) = ubs$ such that
	\[
		q\xtransto{~~~c~~~}
		q'''
		\qquad\text{in }\M
	\]
	and
	\[
			(q,\epsilon)
			\xTransto{u}
			(q,u)
			\xtransto{b}
			(q'',u'')
			\xTransto{s}
			(q''',\epsilon)
		\qquad\text{in }\rho_\S(\M).
	\]
	Since all involved states are reachable, $\R^*(u)$ and $\R^*(s)$ are defined.\twnote{}
	In total, we have that
	\[
		(\R*\S)(c) = \R^*(ubs) = \R^*(u) \, \R(b)\, \R^*(s) = \R^*(u) \, var\,\R^*(s)
	\]
	and in particular
	\[
		\R^*(u)\, va \le (\R*\S)(c).
	\]
	Thus, the state
	\[
		h((q,u),v) = (q, \R^*(u)\,v)\text{ in }\rho_{\R*\S}(\M)
	\]
	has an $a$-transition to
	\[
		h((q',u'),v') = (q', \R^*(u')\,v'),
	\]
	as desired.

	For the verification that $h$ is a simulation in the converse direction,
	i.e.~from $\rho_{\R*\S}(\M)$ to $\rho_\R(\rho_\S(\M))$, consider a transition
	\[
		(q,u) \xtransto{a} (q',u')
		\qquad
		\text{in }\rho_{\R*\S}(\M).
	\]
	Again, using \autoref{rho transition}, we obtain $c\in C$ with
	\[
		(\R*\S)(c) = uav
		\text{ and }
		q \xtransto{c} q''
		\text{ and }
		(q',u') \xTransto{v} (q'',\epsilon).
	\]
	By the definition of $\R*\S$, we thus obtain that $u\in A^*$ must be of the shape
	\[
		\R^*(w)\,r = u
		\qquad\text{for some }w\in \dom(\R)^*
	\]
	By the definition of $\R*\S$, we have
	\[
		w\,b\le \S(c)
		\text{ with }
		u = \R^*(w)\,r
		\text{ and } ra \le \R(b).
	\]
	Then, $h((q,w),r) = (q,u)$ and we distinguish:\twnote{}
	\begin{itemize}
	\item If $ua = (\R*\S)(c)$, then the above $a$-transition in $\rho_{\R^*\S}(\M)$ is produced by rule \eqref{rho end}, and we can use the same rule in $\rho_{\S}(\M)$ and $\rho_\R(\rho_{\S}(\M))$ to establish the desired transition $a$-transition to
		\[
			((q,w), r) \xtransto{a}
			((q,\epsilon), \epsilon)
			\quad\text{in }\rho_{\R}(\rho_{\S}(\M)).
		\]
	\item If $ua \lneqq (\R*\S)(c)$ but $ra = \R(b)$,
		then we use the rule \eqref{rho mid} in $\rho_{\S}(\M)$ and but rule \eqref{rho end} in $\rho_\R(\rho_{\S}(\M))$:
		\[
			((q,w), r) \xtransto{a}
			((q,w\,b), \epsilon)
			\quad\text{in }\rho_{\R}(\rho_{\S}(\M)).
		\]
	\item If $ua \lneqq (\R*\S)(c)$ and $ra \lneqq \R(b)$,
		then we use rules \eqref{rho mid} in both $\rho_{\S}(\M)$ and $\rho_\R(\rho_{\S}(\M))$:
		\[
			((q,w), r) \xtransto{a}
			((q,w), r\,a)
			\quad\text{in }\rho_{\R}(\rho_{\S}(\M)).
		\]
	\end{itemize}
	Hence, in any of the above cases, we have a corresponding $a$-transition in 
	$\rho_{\R}(\rho_{\S}(\M))$.
\end{proofappendix}

\begin{proofappendix}[Details for]{concretization no code composition}
\label{details contraction composition}
We fix $\SI$ to be the identity relation for both codes, and hence omit $\SI$
in the following.
It is a standard result about Galois connections (and adjunctions in general) that they are compatible with composition: the right-adjoint of the composition of two functions is equal to the composition of the respective right-adjoints. One only needs to be warned that \textqt{equal} here refers to the equality induced by the order $\sqsubseteq$, which means mutual simulation:
For all action codes $\R\in \Code(A,B)$, $\S\in\Code(B,C)$ and $\M\in \LTS(C)$:
\[
	\gamma_{\R*\S}(\M)
	\sqsubseteq
	\gamma_{\R}(\gamma_\S(\M))
	\qquad
	\text{and}
	\qquad
	\gamma_{\R*\S}(\M)
	\sqsupseteq
	\gamma_{\R}(\gamma_\S(\M)).
\]
This is however weaker than the notion of isomorphism we consider (\autoref{def:isomorphism}). Concretely, we even have the following counterexample with
\(
	\gamma_{\R*\S}(\M)
	\not\cong
	\gamma_{\R}(\gamma_\S(\M))
\).
	Consider sets $A =
	\set{a}$, $B= \set{b}$, $C = \set{c}$ and the action codes
	\[
		\begin{array}{lll}
		\R\in \Code(A,B)
		& \R\colon B\partialto A^+
		& b \mapsto a\,a
		\\
		\S\in \Code(B,C)
		& \S\colon C\partialto B^+
		& \text{undefined everywhere}
		\end{array}
	\]
	Start with a singleton system that has no transitions:
	\[
		\M :=
		\begin{tikzpicture}[lts,baseline=(q0.base)]
			\node[initial,state] (q0) {\(q_0\)};
			\drawLtsFrame
		\end{tikzpicture}
		\qquad\text{ in }\LTS(C)
	\]
	For the empty $\S$, the concretization $\gamma_\S\colon \LTS(C)\to \LTS(B)$
	sends this into the system
	\[
		\gamma_\S(\M) =
		\begin{tikzpicture}[lts,baseline=(q0.base),x=2cm]
			\node[initial,state] (q0) {\(q_0,\epsilon\)};
			\node[state] (chi) at (1,0) {\(\chi\)};
			\draw[transition] (q0) edge node {\(b\)} (chi);
			\draw[transition] (chi) edge[loop right] node {\(b\)} (chi);
			\drawLtsFrame
		\end{tikzpicture}
		\qquad\text{ in }\LTS(B)
	\]
	We have a $b$-transition from $q_0$ to $\chi$ because $\S(b)$ is undefined.
	For the next action code $\R$, the concretization $\gamma_\R\colon \LTS(B)\to
	\LTS(A)$ treats $\chi$ as an ordinary state, so it produces the following LTS, in which we have omitted the unreachable sink state introduced by $\gamma_\R$:
	\[
		\gamma_\R(\gamma_\S(\M)) \cong 
		\begin{tikzpicture}[lts,baseline=(q0.base),x=2cm]
			\node[initial,state] (q0) at (-0.5,0) {\((q_0,\epsilon),\epsilon\)};
			\node[state] (q0a) at (1,0) {\((q_0,\epsilon),a\)};
			\node[state] (chi) at (2.5,0) {\(\chi,\epsilon\)};
			\node[state] (chia) at (3.5,0) {\(\chi,a\)};
			\draw[transition] (q0) edge node {\(a\)} (q0a);
			\draw[transition] (q0a) edge node {\(a\)} (chi);
			\draw[transition,bend left] (chi) edge node {\(a\)} (chia);
			\draw[transition,bend left] (chia) edge node {\(a\)} (chi);
			\drawLtsFrame
		\end{tikzpicture}
		\quad\text{ in }\LTS(A)
	\]
	Here, we omitted the unreachable chaos state $\chi$ introduced by
	$\gamma_\R$, because the unreachable parts are not relevant for our notion of
	isomorphism $\cong$.

	On the other hand, the composed action code $\R*\S\in \Code(A,C)$ is
	undefined everywhere, so analogously to $\gamma_\S$, concretization for
	$\R*\S$ sends above $\M\in \LTS(C)$ to
	\[
		\gamma_{\R*\S}(\M) =
		\begin{tikzpicture}[lts,baseline=(q0.base),x=2cm]
			\node[initial,state] (q0) {\(q_0,\epsilon\)};
			\node[state] (chi) at (1,0) {\(\chi\)};
			\draw[transition] (q0) edge node {\(a\)} (chi);
			\draw[transition] (chi) edge[loop right] node {\(a\)} (chi);
			\drawLtsFrame
		\end{tikzpicture}
		\quad\text{ in }\LTS(A).
	\]
	Obviously, $\gamma_{\R*\S}(\M)$ is not isomorphic to
	$\gamma_{\R}(\gamma_{\S}(\M))$, but there are canonical simulations in either
	direction, induced by the Galois connection between $\alpha$ and $\gamma$,
	using that $\alpha$ commutes with action code composition.
\end{proofappendix}

\begin{proofappendix}{alpha implementation}
	The notion of delay simulation equivalence that is used in the statement of the theorem can be formally defined as follows:
	
	\begin{definition}
		Let $\M = \tuple{Q,q_0,\transto} \in \LTS(A) $ be an LTS, where $A$ is a set of labels that contains the \emph{hidden action} $\tau$.
		Let $q\xTransto{~~} q'$ denote that there is finite sequence of states
		$r_0,\ldots,r_n\in Q$ such that $r_0 = q$, $r_n=q'$ and $r_{i-1}\xtransto{\tau} r_i$ for all $1\le i\le n$. A relation $S \subseteq Q \times Q$ is called a \emph{delay simulation} if it satisfies the following transfer property:
		\begin{itemize}
			\item 
			If $(q, r) \in S$ and $q \xtransto{a} q'$ then either $a = \tau$ and $q = q'$, or 
			$\exists r', r''$ such that $r \xTransto{~~} r' \xtransto{a} r''$ and $(q', r'') \in S$.
		\end{itemize}
		We write $q \sqsubseteq_d r$ if there exists a weak delay simulation that relates $q$ and $r$.
		We say that $q$ and $r$ are \emph{delay simulation equivalent}, notation $q \equiv_d r$, if both $q \sqsubseteq_d r$ and $r \sqsubseteq_d q$.
	\end{definition}
	We describe the behavior of an implementation for $\M$ and an adaptor for $\R$ formally as expressions in Milner's Calculus of Communicating Systems (CCS) \cite{Mi89}. 
	The semantics of CCS is defined in terms of an infinite LTS in which the states are CCS expressions, and the transitions between states are defined by structural operational semantics rules given in \cite{Mi89}.
	In the rest of this proof, we will assume that the reader is familiar with the CCS calculus.
	In our CCS expressions we use action names taken from $I$, $O$, $X$ and $Y$, and without loss of generality we assume these four sets to be disjoint.
	Process ${\sf Impl}(\M)$ describes the behavior of an implementation for $\M$ in which inputs and outputs are separated and occur sequentially. We define
	 ${\sf Impl}(\M)$ as the CCS expression $M(q_0^\M)$, where for $q \in Q^\M$ and $i \in I$,
	\begin{eqnarray*}
		M(q) & = & \sum_{i \in I} i \cdot M(q,i)\\
		M(q, i) & = & \sum_{o \in O, q' \in Q^\M ~ \mid ~ q \xtransto{i/o} q'} \bar{o} \cdot M(q')
	\end{eqnarray*}
	Similarly, we introduce a process ${\sf Adaptor}(\R)$ that describes the behavior of an adaptor for action code $\R$.  Following the pseudocode of Algorithm~\ref{alg: adaptor}, we define ${\sf Adaptor}(\R)$ as the CCS expression $P(r_0)$,
	where for $r \in R$ and $x \in X$,
	\begin{eqnarray*}
		P(r) & = & \sum_{x \in X} x \cdot Q(r, x)\\
		Q(r, x) & = & \bar{i} \cdot R(r, x) \hspace{2cm}  \mbox{if } r \mbox{ is internal and } i \mbox{ winning for } x \mbox{ in } r\\
		R(r, x) & = & \sum_{o \in O, r' \in R ~ \mid ~ \xtransto{i/o}_\M \wedge r \xtransto{i/o} r' \wedge i \mbox{ winning for } x \mbox{ in } r} o \cdot Q(r',x)\\
		Q(r, x) & = & \overline{\pi_2 ( l(r))} \cdot P(r_0) \hspace{1cm} \mbox{if } r \mbox{ is a leaf}
	\end{eqnarray*}
	Processes ${\sf Adaptor}(\R)$ and ${\sf Impl}(\M)$ may synchronize via actions taken from $I\cup O $.
	If we compose these processes using the CCS composition operator $\mid$,
	and apply the CCS restriction operator $\setminus$ to hide all communications,
	we obtain a CCS expression that describes the behavior of the parallel composition of the adaptor and the SUT.
	 We claim that this composition is delay simulation equivalent to the expression ${\sf Impl}(\alpha_\R(\M))$ that describes the behavior of an implementation of $\alpha_\R(\M)$:
	 \begin{eqnarray}
	 \label{delay simulation equation}
	  ({\sf Adaptor}(\R) \mid {\sf Impl}(\M)) \setminus (I\cup O) & \equiv_d & {\sf Impl}(\alpha_\R(\M))
	 \end{eqnarray}
	 Here we define
	 ${\sf Impl}(\alpha_\R(\M))$ as the CCS expression $N(q_0^{\alpha(\M)})$, where for $q \in Q^{\alpha(\M)}$ and $x \in X$,
	 \begin{eqnarray*}
	 	N(q) & = & \sum_{x \in X} x \cdot N(q,x)\\
	 	N(q, x) & = & \sum_{y \in Y, q' \in Q^{\alpha(\M)} ~ \mid ~ q \xtransto{x/y}_{\alpha(M)} q'} \bar{y} \cdot N(q')
	 \end{eqnarray*}
 	Consider the following relation $S$ between CCS expressions:
 	\begin{eqnarray*}
 		S & = & \{  ((P(r_0) \mid M(q))\setminus (I\cup O),~ N(q)) ~\mid ~ q \in Q^{\alpha(\M)} \} \\
 		& & \cup ~~  \{ ( (Q(r,x) \mid M(q'))\setminus (I\cup O), N(q, x) ) ~\mid ~ q \in Q^{\alpha(\M)},~ q'\in Q^\M, \\
 		& & \hspace{1 cm} ~r \in R \mbox{ winning for } x \in X ~\wedge ~ \exists \sigma \in (I \times O)^* \colon  r_0 \xTransto{\sigma}_\R r \wedge q \xTransto{\sigma}_\M q' \}\\
 		& & \cup ~~  \{ ( (R(r,x) \mid M(q', i))\setminus (I\cup O),~ N(q, x) ) ~\mid ~ q \in Q^{\alpha(M)},~ q'\in Q^\M, \\
 		& & \hspace{1 cm} ~r \in R \mbox{ winning for } x \in X \mbox{ with } i \in I ~\wedge ~ \exists \sigma \in (I \times O)^* \colon  r_0 \xTransto{\sigma}_\R r \wedge q \xTransto{\sigma}_\M q' \}
 	\end{eqnarray*}
 We claim that $S$ is a delay simulation relation.
 In order to prove this, we check that the transfer property holds for all pairs of related states and enabled transitions:
 \begin{enumerate}
 	\item 
 	Assume $( (P(r_0) \mid M(q))\setminus (I\cup O),~ N(q)) \in S$ and
 	$(P(r_0) \mid M(q))\setminus (I\cup O) \xtransto{x} (Q(r_0, x) \mid M(q))\setminus (I\cup O)$, for some $x \in X$.
 	We observe that $N(q) \xtransto{x} N(q,x)$ and note that $( (Q(r_0,x) \mid M(q))\setminus (I\cup O),~ N(q, x)) \in S$ since $r_0$ is winning for $x$, $r_0 \xTransto{\sigma}_\R r_0$ and $q \xTransto{\sigma}_\M q$.
 	\item 
 	Assume $( (Q(r,x) \mid M(q'))\setminus (I\cup O),~ N(q, x)) \in S$ and
 	$(Q(r,x) \mid M(q'))\setminus (I\cup O) \xtransto{\tau} (R(r,x) \mid M(q',i)) \setminus (I \cup O)$, for $r$ internal and $i$ the unique input that is winning for $x$ in $r$.
 	By the assumption, $q \in Q^{\alpha(M)}$, $q'\in Q^\M$, and there exists
 	$\sigma \in (I \times O)^*$ such that $r_0 \xTransto{\sigma}_\R r$ and $q \xTransto{\sigma}_\M q'$. Then $( (R(r,x) \mid M(q', i))\setminus (I\cup O),~ N(q, x)) \in S$, as required.
 	\item 
 	Assume $((R(r,x) \mid M(q', i))\setminus (I\cup O),~ N(q, x)) \in S$ and
 	$(R(r,x) \mid M(q', i))\setminus (I\cup O) \xtransto{\tau} (Q(r', x) \mid M(q'')) \setminus (I \cup O)$, with $r \xtransto{i/o} r'$ and $q'\xtransto{i/o} q''$.
 	By the assumption, $q \in Q^{\alpha(M)}$, $q'\in Q^\M$, $r$ is winning for $x$ with $i$, and there exists
 	$\sigma \in (I \times O)^*$ such that $r_0 \xTransto{\sigma}_\R r$ and $q \xTransto{\sigma}_\M q'$. Then $q''\in Q^\M$ and,
 	by definition of winning, $r'$ is winning for $x$.
 	Moreover, if we take $\sigma' = \sigma \cdot (i,o)$, then $r_0 \xTransto{\sigma'}_\R r'$ and $q \xTransto{\sigma'}_\M q''$. This implies that
 	$( (Q(r', x) \mid M(q''))\setminus (I\cup O),~ N(q, x)) \in S$, as required.
 	\item 
 	Assume $( (Q(r,x) \mid M(q'))\setminus (I\cup O),~ N(q, x)) \in S$ and
 	$(Q(r,x) \mid M(q'))\setminus (I\cup O) \xtransto{\overline{\pi_2 ( l(r))}} (P(r_0) \mid M(q')) \setminus (I \cup O)$, for $r$ a leaf.
 	By the assumption, $q \in Q^{\alpha(\M)}$, $q'\in Q^\M$, $r$ is winning for $x$, and there exists
 	$\sigma \in (I \times O)^*$ such that $r_0 \xTransto{\sigma}_\R r$ and $q \xTransto{\sigma}_\M q'$. 
 	By definition of the contraction operator, $q \xtransto{l(r)}_{\alpha(\M)} q'$ and $q' \in Q^{\alpha(M)}$.
 	But this means $N(q,x) \xtransto{\overline{\pi_2 ( l(r))}} N(q')$.
 	Now observe that $((P(r_0) \mid M(q'))\setminus (I\cup O),~ N(q')) \in S$, as required.	
 \end{enumerate}
Next consider the following relation $T$ between CCS expressions:
\begin{eqnarray*}
	T & = & \{  (N(q),~ (P(r_0) \mid M(q))\setminus (I\cup O) ) ~\mid ~ q \in Q^{\alpha(\M)} \} \\
	& & \cup ~~  \{ (N(q, x),~ (Q(r_0, x) \mid M(q))\setminus (I\cup O)  ) ~\mid ~ q \in Q^{\alpha(\M)}  \}
\end{eqnarray*}
We claim that $T$ is a delay simulation relation, and
check that the transfer property holds for all pairs of related states and enabled transitions:
\begin{enumerate}
\item 
Assume $(N(q),~ (P(r_0) \mid M(q))\setminus (I\cup O)) \in T$ and $N(q) \xtransto{x} N(q,x)$, for some $x \in X$.
We observe that
$(P(r_0) \mid M(q))\setminus (I\cup O) \xtransto{x} (Q(r_0, x) \mid M(q))\setminus (I\cup O)$ and note that $(N(q, x),~  (Q(r_0,x) \mid M(q))\setminus (I\cup O)) \in T$. 
\item 
Assume $(N(q, x),~  (Q(r_0,x) \mid M(q))\setminus (I\cup O)) \in T$ and
$N(q, x) \xtransto{\bar{y}} N(q')$.
Then $\alpha(\M)$ has a transition $q \xtransto{x/y} q'$ and $q' \in Q^{\alpha(\M)}$.
By definition of $\alpha(\M)$, $\R$ has a leaf $r$ with $l(r) = (x, y)$ and there exists a sequence $\sigma$ such that $r_0 \xTransto{\sigma}_\R r$ and $q \xTransto{\sigma}_\M q'$. 	
Let
\begin{eqnarray*}
	\sigma & = & (i_1, o_1) (i_2, o_2) \cdots (i_n, o_n)
\end{eqnarray*}
Then $\R$ has states $r_1 ,\ldots, r_n$ and $\M$ has states $s_0 ,\ldots, s_n$ such that:
\[
r_0 \xtransto{i_1/o_1}_\R r_1 \xtransto{i_2/o_2}_\R r_2 \cdots \xtransto{i_n/o_n}_\R r_n
\]
\[
q = s_0 \xtransto{i_1/o_1}_\M s_1 \xtransto{i_2/o_2}_\M s_2 \cdots \xtransto{i_n/o_n}_\M s_n = q'
\]
From these runs in $\R$ and $\M$ we may construct a sequence of $\tau$-transitions:
\begin{eqnarray*}
	(Q(r_0, x) \mid M(s_0))\setminus (I\cup O) & \xtransto{\tau} & (R(r_0, x) \mid M(s_0, i_1))\setminus (I\cup O)\\
	& \xtransto{\tau} & (Q(r_1, x) \mid M(s_1))\setminus (I\cup O)\\
	& & \vdots\\
	& \xtransto{\tau} & (R(r_{n-1}, x) \mid M(s_{n-1}, i_n))\setminus (I\cup O)\\
	& \xtransto{\tau} & (Q(r_n, x) \mid M(s_n))\setminus (I\cup O)
\end{eqnarray*}
Note that our assumptions that $\R$ is determinate and has a winning strategy for every input $x \in X$ impy that the inputs that occur in $\sigma$ are always winning.
From the above sequence of $\tau$-transitions we may conclude
\[
(Q(r_0, x) \mid M(q))\setminus (I\cup O) ~~  \xTransto{~~} ~~ (Q(r, x) \mid M(q'))\setminus (I\cup O)
\]
Since $(Q(r, x) \mid M(q'))\setminus (I\cup O) \xtransto{\bar{y}} (P(r_0) \mid M(q'))\setminus (I\cup O)$ and $(N(q'),~ (P(r_0) \mid M(q'))\setminus (I\cup O) ) \in T$, the transfer property follows.
\end{enumerate}
Because $S$ is a delay simulation from $ ({\sf Adaptor}(\R) \mid {\sf Impl}(\M)) \setminus (I\cup O)$ to ${\sf Impl}(\alpha_\R(\M))$, and $T$ is a delay simulation from ${\sf Impl}(\alpha_\R(\M))$ to $({\sf Adaptor}(\R) \mid {\sf Impl}(\M)) \setminus (I\cup O)$,  identity (\ref{delay simulation equation}) follows, and thereby the theorem.
\end{proofappendix}

\begin{proofappendix}{La preservation properties by contraction}
		Assume $\M$ is output deterministic and $\R$ is determinate.  Let $\N = \alpha_\R(\M)$.
		Suppose that $\N$ has transitions $q \xtransto[\N]{x/y'} q'$ and $q \xtransto[\N]{x/y''} q''$. We need to show $y' = y''$ and $q' = q''$.
		The transitions have been derived using rule (\ref{Rule A2}), and
		formulated for tree-shaped action codes $\R$, we know there are generalized
		transitions
		\begin{eqnarray*}
			q \xTransto[\M]{u'/s'} q' & r_0 \xTransto[\R]{u'/s'} r' & r'\in L ~~ l(r') = (x, y')\\
			q \xTransto[\M]{u''/s''} q'' & r_0 \xTransto[\R]{u''/s''} r'' & r''\in L ~~ l(r'') = (x, y'')
		\end{eqnarray*}
		Now since $\R$ is determinate, the first inputs in $u'$ and $u''$ must be identical.
		But since $\M$ is output deterministic, this implies that the first outputs in $s'$ and $s''$ must also be identical.
		Moreover, the paths from $q$ to $q'$ and $q''$ in $\M$ share the same initial transition.
		Since action codes are deterministic, the paths from $r_0$ to $r'$ and $r''$ in $\R$ also share the same initial transition. By repeating this line of reasoning,
		we can ``zip together'' the paths from $q$ to $q'$ and $q''$ in $\M$, and the paths from $r_0$ to $r'$ and $r''$ in $\R$, and obtain
		$u' = u''$, $s' = s''$, $q' = q''$, $r' = r''$ and $y' = y''$, as required.
\end{proofappendix}

\end{document}